\documentclass[twocolumn]{aastex63}

\hypersetup{linkcolor=red,citecolor=blue,filecolor=cyan,urlcolor=black}

\newcommand{\OIII}{[\ion{O}{3}]}
\newcommand{\OII}{[\ion{O}{2}]}

\newcommand{\CIII}{\ion{C}{3}]}

\newcommand{\CIV}{\ion{C}{4}}
\newcommand{\HeII}{He\,{\scriptsize II}}

\newcommand{\hst}{\textit{HST}}
\newcommand{\spitzer}{\textit{Spitzer}}
\newcommand{\jwst}{\textit{JWST}}

\newcommand{\mirage}{\texttt{MIRAGE}}
\newcommand{\grizli}{\texttt{Grizli}}
\newcommand{\lya}{Ly$\alpha$}
\newcommand{\ergscm}{erg s$^{-1}$ cm$^{-2}$}

\usepackage{relsize}

\received{April 20, 2022}
\revised{??}
\accepted{??}
\submitjournal{ApJ}

\graphicspath{{./}{figures/}}

\shorttitle{The GLASS JWST ERS Program}
\shortauthors{Treu et al.}

\begin{document}

\title{The GLASS James Webb Space Telescope Early Release Science Program. I. Survey Design and Release Plans}


\correspondingauthor{Tommaso Treu}
\email{tt@astro.ucla.edu}


\author[0000-0002-8460-0390]{T. Treu}
\affiliation{Department of Physics and Astronomy, University of California, Los Angeles, 430 Portola Plaza, Los Angeles, CA 90095, USA}

\author[0000-0002-4140-1367]{G. Roberts-Borsani}
\affiliation{Department of Physics and Astronomy, University of California, Los Angeles, 430 Portola Plaza, Los Angeles, CA 90095, USA}

\author[0000-0001-5984-0395]{M. Bradac}
\affiliation{
University of Ljubljana, Department of Mathematics and Physics, Jadranska ulica 19, SI-1000 Ljubljana, Slovenia}
\affiliation{
Department of Physics and Astronomy, University of California Davis, 1 Shields Avenue, Davis, CA 95616, USA}

\author[0000-0003-2680-005X]{G. Brammer}
\affiliation{Cosmic Dawn Center (DAWN), Denmark}
\affiliation{Niels Bohr Institute, University of Copenhagen, Jagtvej 128, DK-2200 Copenhagen N, Denmark}

\author[0000-0003-3820-2823]{A. Fontana}
\affiliation{INAF Osservatorio Astronomico di Roma, Via Frascati 33, 00078 Monteporzio Catone, Rome, Italy}

\author[0000-0002-6586-4446]{A. Henry}
\affiliation{Space Telescope Science Institute, 3700 San Martin Drive, Baltimore MD, 21218} 
\affiliation{Center for Astrophysical Sciences, Department of Physics and Astronomy, Johns Hopkins University, Baltimore, MD, 21218} 

\author[0000-0002-3407-1785]{C. Mason}
\affiliation{Cosmic Dawn Center (DAWN), Denmark}
\affiliation{Niels Bohr Institute, University of Copenhagen, Jagtvej 128, DK-2200 Copenhagen N, Denmark}

\author[0000-0002-8512-1404]{T. Morishita}
\affiliation{Infrared Processing and Analysis Center, Caltech, 1200 E. California Blvd., Pasadena, CA 91125, USA}

\author[0000-0001-8940-6768 ]{L. Pentericci}
\affiliation{INAF Osservatorio Astronomico di Roma, Via Frascati 33, 00078 Monteporzio Catone, Rome, Italy}

\author[0000-0002-9373-3865]{X. Wang}
\affiliation{Infrared Processing and Analysis Center, Caltech, 1200 E. California Blvd., Pasadena, CA 91125, USA}

\author{A.~Acebron}
\affiliation{Dipartimento di Fisica, Università degli Studi di Milano, Via Celoria 16, I-20133 Milano, Italy}
\affiliation{INAF - IASF Milano, via A. Corti 12, I-20133 Milano, Italy}

\author[0000-0002-9921-9218]{M.~Bagley}
\affiliation{Department of Astronomy, The University of Texas at Austin, Austin, TX, USA}

\author[0000-0003-1383-9414]{P.~Bergamini}
\affiliation{Dipartimento di Fisica, Università degli Studi di Milano, Via Celoria 16, I-20133 Milano, Italy}
\affiliation{INAF - OAS, Osservatorio di Astrofisica e Scienza dello Spazio di Bologna, via Gobetti 93/3, I-40129 Bologna, Italy}

\author{D.~Belfiori}
\affiliation{INAF Osservatorio Astronomico di Roma, Via Frascati 33, 00078 Monteporzio Catone, Rome, Italy}

\author{A.~Bonchi}
\affiliation{INAF Osservatorio Astronomico di Roma, Via Frascati 33, 00078 Monteporzio Catone, Rome, Italy}
\affiliation{ASI-Space Science Data Center,  Via del Politecnico, I-00133 Roma, Italy}

\author[0000-0003-4109-304X]{K.~Boyett}
\affiliation{School of Physics, University of Melbourne, Parkville 3010, VIC, Australia}
\affiliation{ARC Centre of Excellence for All Sky Astrophysics in 3 Dimensions (ASTRO 3D), Australia}
\author[ 0000-0003-4432-5037]{K.~Boutsia}
\affiliation{ Carnegie Observatories, Las Campanas Observatory, Casilla 601, La Serena, Chile}

\author[0000-0003-2536-1614]{A. Calabr\'o}
\affiliation{INAF Osservatorio Astronomico di Roma, Via Frascati 33, 00078 Monteporzio Catone, Rome, Italy}

\author[0000-0001-6052-3274]{G.~B.~Caminha}
\affiliation{Max-Planck-Institut f\"ur Astrophysik, Karl-Schwarzschild-Str. 1, D-85748 Garching, Germany}
\author[0000-0001-9875-8263]{M.~Castellano}
\affiliation{INAF Osservatorio Astronomico di Roma, Via Frascati 33, 00078 Monteporzio Catone, Rome, Italy}

\author[0000-0002-6317-0037]{A.~Dressler}
\affiliation{The Observatories, The Carnegie Institution for Science, 813 Santa Barbara St., Pasadena, CA 91101, USA}
\author[0000-0002-3254-9044]{K. Glazebrook}
\affiliation{Centre for Astrophysics and Supercomputing, Swinburne University of Technology, PO Box 218, Hawthorn, VIC 3122, Australia}
\affiliation{ARC Centre of Excellence for All Sky Astrophysics in 3 Dimensions (ASTRO 3D), Australia)}
\author[0000-0002-5926-7143]{C. Grillo}
\affiliation{Dipartimento di Fisica, Università degli Studi di Milano, Via Celoria 16, I-20133 Milano, Italy}
\affiliation{INAF - IASF Milano, via A. Corti 12, I-20133 Milano, Italy}
\author[0000-0003-4239-4055]{C. Jacobs}
\affiliation{Centre for Astrophysics and Supercomputing, Swinburne University of Technology, PO Box 218, Hawthorn, VIC 3122, Australia}
\affiliation{ARC Centre of Excellence for All Sky Astrophysics in 3 Dimensions (ASTRO 3D), Australia)}
\author[0000-0001-5860-3419]{T. Jones}
\affiliation{Department of Physics and Astronomy, University of California Davis, 1 Shields Avenue, Davis, CA 95616, USA}
\author[0000-0003-3142-997X]{P. L. Kelly}
\affiliation{Minnesota Institute for Astrophysics, University of Minnesota, 116 Church St. SE, Minneapolis, MN 55455 USA}

\author[0000-0003-4570-3159]{N. Leethochawalit}
\affiliation{School of Physics, University of Melbourne, Parkville 3010, VIC, Australia}
\affiliation{ARC Centre of Excellence for All Sky Astrophysics in 3 Dimensions (ASTRO 3D), Australia}

\author[0000-0001-6919-1237]{M.A. Malkan}
\affiliation{Department of Physics and Astronomy, University of California, Los Angeles, 430 Portola Plaza, Los Angeles, CA 90095, USA}

\author[0000-0001-9002-3502]{D.~Marchesini}
\affiliation{
Department of Physics and Astronomy, Tufts University, 574 Boston Ave., Medford, MA 02155, USA}
\author[0000-0002-9572-7813]{S.~Mascia}
\affiliation{INAF Osservatorio Astronomico di Roma, Via Frascati 33, 00078 Monteporzio Catone, Rome, Italy}

\author{A.~Mercurio}
\affiliation{INAF – Osservatorio Astronomico di Capodimonte, Via Moiariello 16, I-80131 Napoli, Italy}
\author[0000-0001-6870-8900]{E.~Merlin}
\affiliation{INAF Osservatorio Astronomico di Roma, Via Frascati 33, 00078 Monteporzio Catone, Rome, Italy}

\author[0000-0003-2804-0648 ]{T.~Nanayakkara}
\affiliation{Centre for Astrophysics and Supercomputing, Swinburne University of Technology, PO Box 218, Hawthorn, VIC 3122, Australia}
\affiliation{ARC Centre of Excellence for All Sky Astrophysics in 3 Dimensions (ASTRO 3D), Australia)}

\author[0000-0001-6342-9662]{M.~Nonino}
\affiliation{INAF – Osservatorio Astronomico di Trieste, Via Tiepolo 11, 34143 Trieste, Italy}

\author{D.~Paris}
\affiliation{INAF Osservatorio Astronomico di Roma, Via Frascati 33, 00078 Monteporzio Catone, Rome, Italy}

\author[0000-0001-8751-8360]{B. Poggianti}
\affiliation{INAF Osservatorio Astronomico di Padova, vicolo dell'Osservatorio 5, 35122 Padova, Italy}
\author[0000-0002-6813-0632]{P.~Rosati}
\affiliation{Dipartimento di Fisica e Scienze della Terra, Università degli Studi di Ferrara, Via Saragat 1, I-44122 Ferrara, Italy}
\affiliation{INAF - OAS, Osservatorio di Astrofisica e Scienza dello Spazio di Bologna, via Gobetti 93/3, I-40129 Bologna, Italy}
\author[0000-0002-9334-8705]{P.~Santini}
\affiliation{INAF Osservatorio Astronomico di Roma, Via Frascati 33, 00078 Monteporzio Catone, Rome, Italy}

\author[0000-0002-9136-8876]{C.~Scarlata}\affiliation{
School of Physics and Astronomy, University of Minnesota, Minneapolis, MN, 55455, USA}
\author{H.~V.~Shipley} \affiliation{
Department of Physics and Astronomy, Tufts University, 574 Boston Ave., Medford, MA 02155, USA}
\author{V.~Strait}
\affiliation{Cosmic Dawn Center (DAWN), Denmark}
\affiliation{Niels Bohr Institute, University of Copenhagen, Jagtvej 128, DK-2200 Copenhagen N, Denmark}

\author[0000-0001-9391-305X]{M. Trenti}
\affiliation{School of Physics, University of Melbourne, Parkville 3010, VIC, Australia}
\affiliation{ARC Centre of Excellence for All Sky Astrophysics in 3 Dimensions (ASTRO 3D), Australia}

\author{C.~Tubthong}
\affiliation{
Department of Physics and Astronomy, Tufts University, 574 Boston Ave., Medford, MA 02155, USA}

\author[0000-0002-5057-135X]{E. Vanzella}
\affiliation{INAF - OAS, Osservatorio di Astrofisica e Scienza dello Spazio di Bologna, via Gobetti 93/3, I-40129 Bologna, Italy}
\author[0000-0003-0980-1499]{B. Vulcani}
\affiliation{INAF Osservatorio Astronomico di Padova, vicolo dell'Osservatorio 5, 35122 Padova, Italy}

\author[0000-0002-8434-880X]{L.~Yang}
\affiliation{Kavli Institute for the Physics and Mathematics of the Universe, The University of Tokyo, Kashiwa, Japan 277-8583}

\begin{abstract}
The GLASS James Webb Space Telescope Early Release Science (hereafter GLASS-JWST-ERS) Program will obtain and make publicly available the deepest extragalactic data of the ERS campaign. It is primarily designed to address two key science questions, namely, ``what sources ionized the universe and when?'' and ``how do baryons cycle through galaxies?'', while also enabling a broad variety of first look scientific investigations. In primary mode, it will obtain NIRISS and NIRSpec spectroscopy of galaxies lensed by the foreground Hubble Frontier Field cluster, Abell 2744. In parallel, it will use NIRCam to observe two fields that are offset from the cluster center, where lensing magnification is negligible, and which can thus be effectively considered blank fields. In order to prepare the community for access to this unprecedented data, we describe the scientific rationale, the survey design (including target selection and observational setups), and present pre-commissioning estimates of the expected sensitivity. In addition, we describe the planned public releases of high-level data products, for use by the wider astronomical community.
\end{abstract}

\keywords{editorials, notices --- 
miscellaneous --- catalogs --- surveys}

\section{Introduction}

Compared to all previous facilities, the James Webb Space Telescope's (\jwst) capabilities are unprecedented.
Its spectroscopic and imaging instruments will provide data of the kind beyond those yet seen by any astronomer. However, our community will have to learn rapidly how to obtain and analyze such transformative data most efficiently. The \jwst\ Early Release Science Program aims to quickly provide  public data sets that will enable early scientific investigations as well as answers to practical questions such as: What is the best instrument to use given partially overlapping capabilities? What systematics come with using a $0\farcs2$ slit on extended sources (e.g., galaxies), or a slitless grism on blended narrow lines? What do the optical spectra and images of the highest redshift galaxies look like?

\begin{figure*}
\includegraphics[angle=0,width=\linewidth]{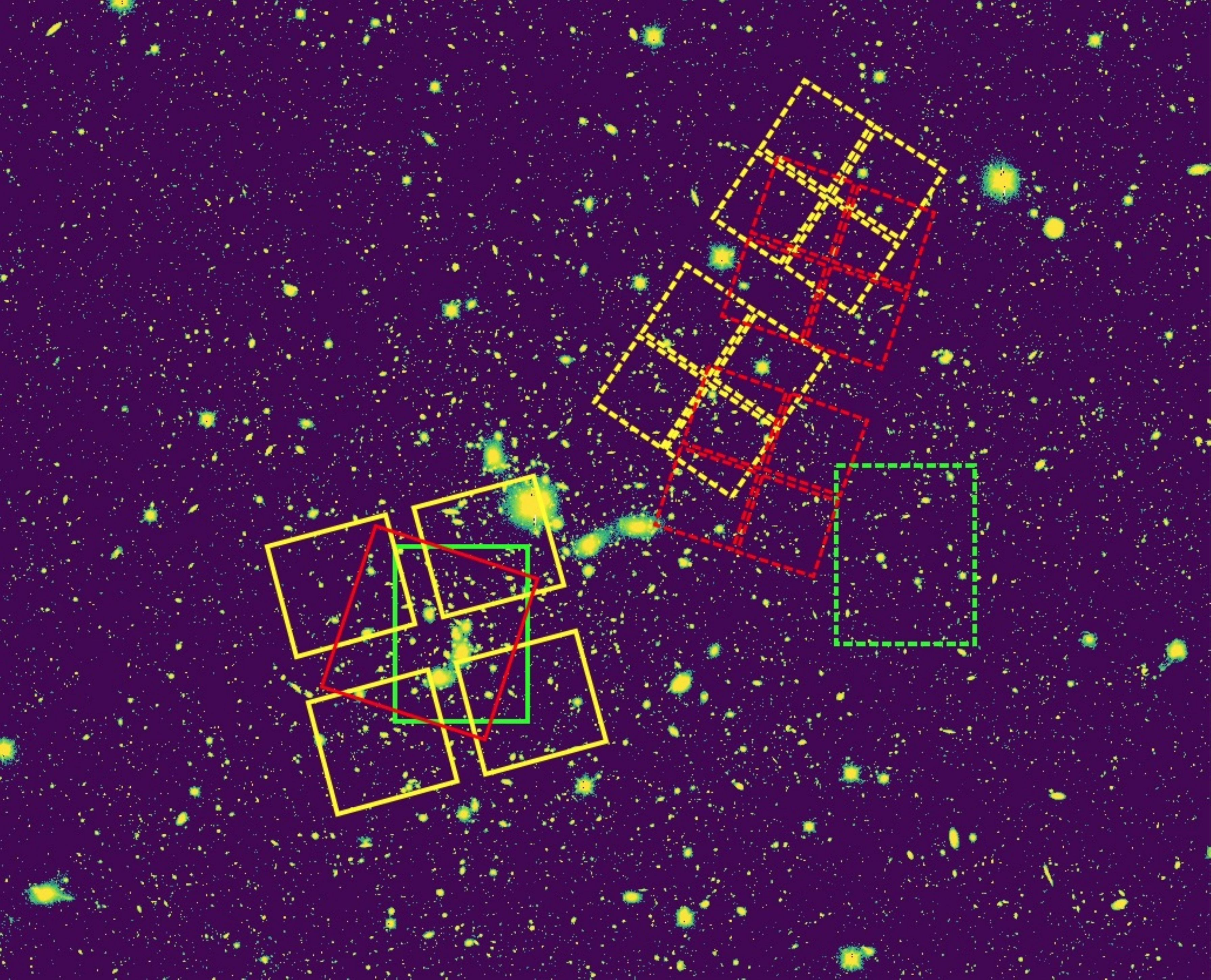}
\caption{\label{fig:FOV} Layout of the GLASS-JWST-ERS observing program. The position of the NIRISS (red solid line) and NIRSpec (yellow solid line) primary fields are shown, along with the NIRCam parallel pointings (dashed lines, color corresponding to the relevant primary instrument) and the extant HFF central (green solid line) and parallel (green dashed line). The background image has been obtained with the Magellan Telescope and will be released as part of the high level data products.}
\end{figure*}

With the above in mind, the GLASS-JWST-ERS program (JWST-ERS-1324: PI Treu) will obtain the deepest ERS observations with NIRISS \citep{NIRISS}, NIRSpec \citep{NIRSPEC1}, and NIRCam \citep{NIRCAM}. In primary mode, it will obtain NIRISS and NIRSpec spectroscopy of galaxies in the Hubble Frontier Field (HFF) cluster, Abell 2744 (A2744). By combining the power of \jwst\ with the power of lensing magnification, it will enable breakthroughs in a broad range of science topics of interest to a large fraction of the extragalactic/high$-z$ astronomical community. Beyond revealing the properties of distant galaxies in unprecedented detail, the unique setup of these observations will contextualize inferences from NIRSpec slits within the 2D picture provided by NIRISS, shedding light on how slit placement and losses bias physical inferences. Simultaneously, the same data will reveal how NIRISS’ spectral resolution affects our understanding of detailed extragalactic physics. In parallel, GLASS-JWST-ERS will obtain NIRCam imaging in two deep fields, with filters and depths designed to identify $z>7$ galaxies. At around one virial radius from the cluster center, these will be approximate blank fields with negligible magnification from lensing. The layout of the fields is shown in Figure~\ref{fig:FOV}.

Here, we present an overview of GLASS-JWST-ERS with the goal of providing the community with the necessary information to take full advantage of the GLASS-JWST-ERS data set and high-level data products. In Section~\ref{sec:drivers} we summarize the key science drivers that were used to design the program and select instrumental setups and observing sequences. In Section~\ref{sec:target} we describe the target selection process, including the selection of the A2744 field and the prioritization of galaxies for NIRSpec observations. In Section~\ref{sec:NIRISS} we describe the NIRISS observational setup and describe and analyze simulated data sets with the goal of providing sensitivity and contamination estimates sufficiently accurate to plan scientific investigations. In Section~\ref{sec:NIRSPEC} we describe the NIRSpec observational setup and provide estimates of the spectral signal-to-noise ratio expected for representative galaxies in the field. In Section~\ref{sec:NIRCAM} we describe the observational setup used for the NIRCam parallel fields and describe and analyze simulated data sets with the goal of providing sensitivity and contamination estimates sufficiently accurate to plan scientific investigations. Section~\ref{sec:Magellan} describes ancillary ground based imaging data that will be released with the program. Section~\ref{sec:releases} describes the plans for release of reduced data and high level data products. Section~\ref{sec:summary} provides a brief summary. Magnitudes are given in the AB system and a standard cosmology with $\Omega_{\rm m}=0.3$ $\Omega_{\Lambda}=0.7$ and H$_0$=70 km s$^{-1}$ Mpc$^{-1}$ is adopted when necessary.

\section{Science Drivers}
\label{sec:drivers}

The GLASS-JWST-ERS program was conceived to enable a broad range of community explorations of the distant universe, ranging from galaxy cluster science to the high redshift universe. 

Among the many compelling science questions, two were singled out to drive the choice of targets, instrumental setup, and exposure times. We refer to those as Key Science Drivers 1 and 2 and briefly discuss them in two subsections below (\S~\ref{ssec:KSD1} and~\ref{ssec:KSD2}). Section~\ref{ssec:ancillaryscience} lists several of many possible scientific investigations that are enabled by this data set, even though they were not used directly to determine the observational strategy. 

\subsection{Key Science Driver 1: What sources ionized the universe and when?} 
\label{ssec:KSD1}

Multiple lines of evidence indicate that the universe was reionized at $z\gtrsim6$ \citep[e.g.,][]{Planck2020A&A...641A...6P,Mason2018ApJ...856....2M,Mason2019MNRAS.489.2669M,Hoag2019ApJ...878...12H,Davies2018ApJ...864..142D,Greig2019MNRAS.484.5094G,Morales2021ApJ...919..120M,Qin2021MNRAS.506.2390Q}. However, the exact timeline of cosmic reionization has not yet been established and the primary sources that governed the process have not yet been identified. Whether reionization was driven by the more numerous, low mass sources with low ionizing photon escape fractions or rarer, high mass sources with enhanced ionizing escape fractions remains a source of debate \citep[e.g.,][]{Rob++10,Finkelstein2019ApJ...879...36F,Mason2019MNRAS.489.2669M,Naidu2020ApJ...892..109N,Matthee2021arXiv211011967M,Matthee2022}. 

\begin{figure}
\includegraphics[width=\columnwidth]{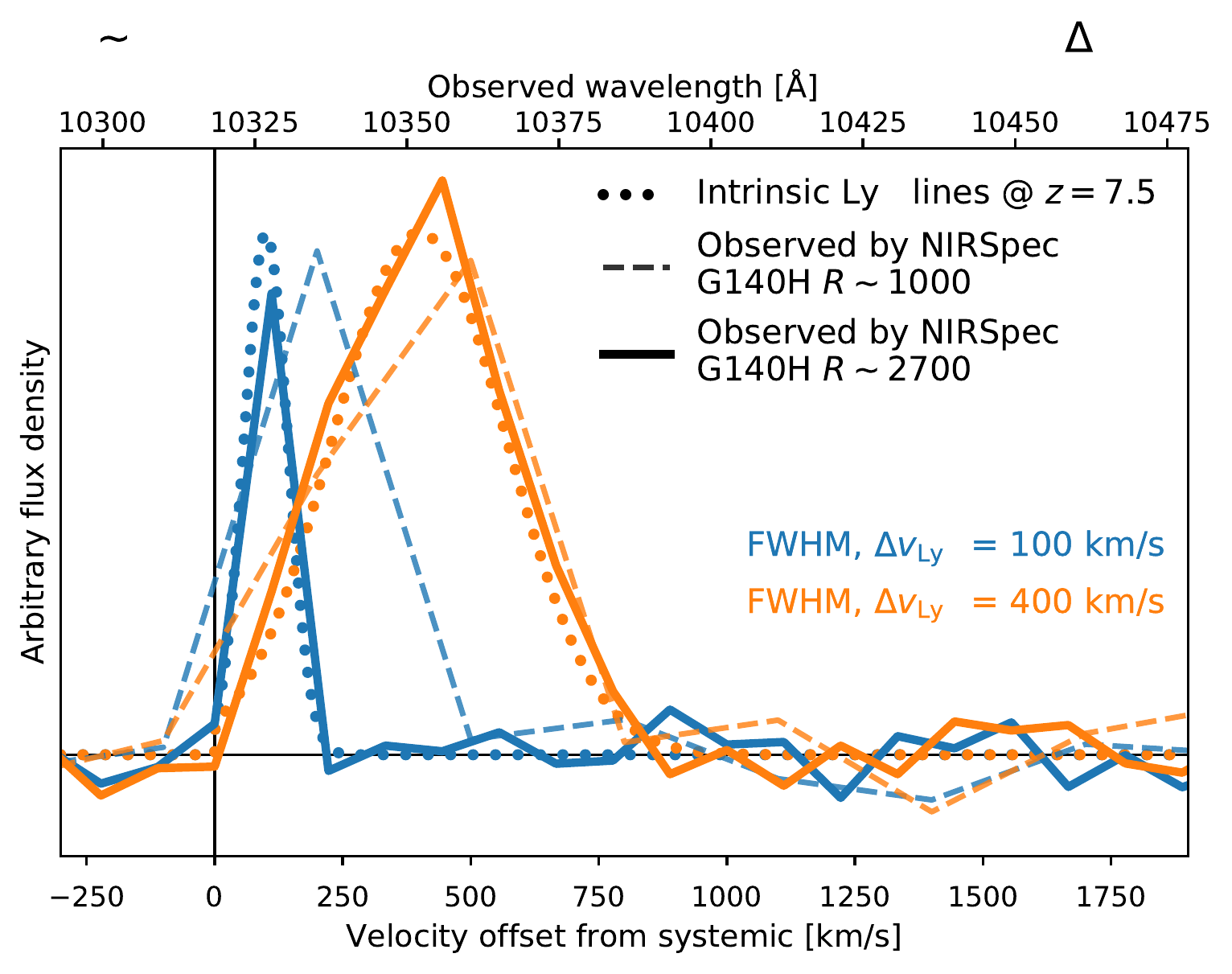}
\caption{\label{fig:NIRSPEC_LYA1} NIRSpec's power to accurately measure \lya\ velocity offsets with respect to systemic. G140H with $R\sim2700$ has a velocity resolution of $\sim100$ km/s, enabling the characterization of the probability of line transmission through the reionizing intergalactic medium (IGM).}
\end{figure}

\begin{figure}
\includegraphics[width=\columnwidth]{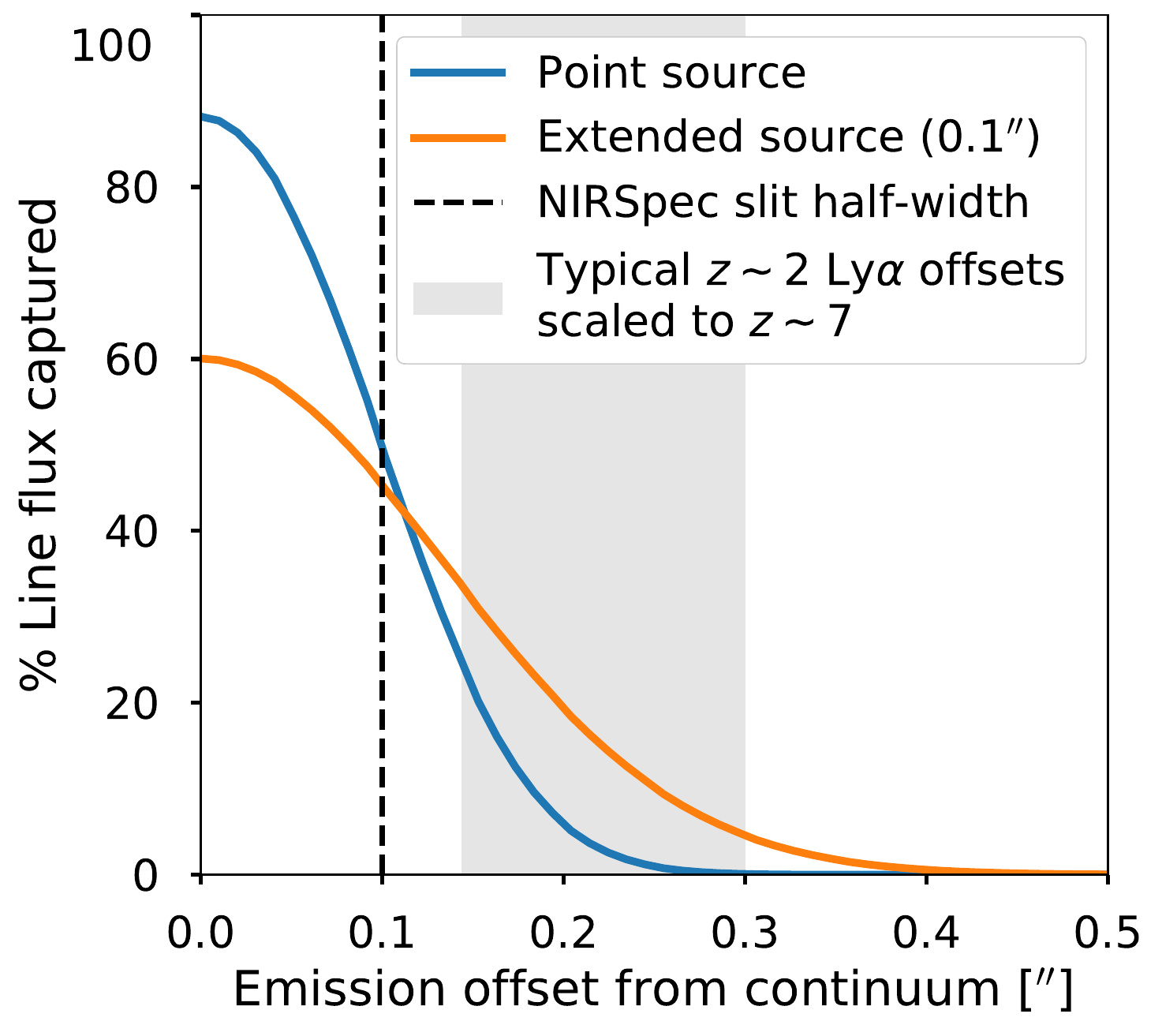}
\caption{\label{fig:NIRSPEC_LYA2} NIRSpec losses for a 
slit positioned on the UV continuum as a function of
  projected offset with \lya, in the idealized case of point-like emission for both. For the $\sim1-3$kpc scale offsets
  typically observed at $z\sim2$ the loss is $\gtrsim$60\%--99\% for a point source.}
\end{figure}

The galaxy UV luminosity function (LF) provides some clues. If ionizing photon escape fractions for Lyman break galaxies are as low as inferred from $z\sim3$ observations \citep[$f_\mathrm{esc} \sim 10\%$, e.g.][]{Steidel2018ApJ...869..123S,Begley2022arXiv220204088B}, providing sufficient photons to ionize the universe requires the UV LF to extend to fainter sources than \hst\ is able to detect in blank fields \citep[e.g.,][though c.f. \citealt{Matthee2021arXiv211011967M}]{Fin++12,Sch++14,Bou++15,MTT15}. HFF data suggest that the LF might continue to rise steeply beyond the blank field limits, and thus testing this hypothesis requires deeper imaging. The intensity of \lya\ detected in Lyman Break Galaxies (LBGs) also yields hints to the progress of reionization: the transmission of \lya\ emission seems to drop at $z\gtrsim6$, suggesting an increasingly neutral intergalactic medium \citep{Fon++10,Tre++13,Sch++14,Pen++14,Mason2018ApJ...856....2M,Mason2019MNRAS.485.3947M,Hoag2019ApJ...878...12H}. 

However, interpreting these data requires an understanding of the interplay between the interstellar and intergalactic media (ISM \& IGM): due to resonant scattering, \lya\ transmission through the IGM depends on the frequency of ISM-escaping photons, which is imprinted by the velocity, density, and spatial distributions of galactic gas \citep[e.g.,][]{Verhamme2006A&A...460..397V,Wofford2013,Henry2015,Rivera-Thorsen2015,Orlitova2018,Hoag2019MNRAS.488..706H,Jaskot2019,Claeyssens2022arXiv220104674C}. Recently, high rates of \lya\ detection in $z > 7.5$ luminous galaxies have been reported \citep{Sta++17}. These rates could be explained if substantial \ion{H}{1} ISM reservoirs in these systems result in \lya\ emission that emerges from the ISM at substantially redshifted velocities  \citep{Mason2018ApJ...857L..11M,Endsley2022arXiv220201219E}.  As such, measuring \lya\ line widths and systemic velocity offsets motivates our requirement for high dispersion rest-frame UV/optical spectra with NIRSpec.  Conversely, measuring the extent of \lya\ requires {\it spatially resolved} 2D spectra, which we will obtain with NIRISS.

GLASS-JWST-ERS aims to pave the way to a new physical understanding of high-$z$ galaxies. In the prime field, the combination of lensing magnification and NIRISS/NIRSpec's spatial and spectral resolution will enable the first studies of how \lya\ propagates through the ISM/IGM at $z\gtrsim7$. In the parallel fields, it will identify an HFF-comparable sample of $7<z<9$ LBGs with NIRCam and more than double the census of currently known galaxies at $z>9$. Using these data, the program will improve dramatically compared to previous estimates with the Spitzer Space Telescope on the rest-frame optical colors, sizes and global properties of galaxies at $z>7$.

\begin{figure*}
\centering
\includegraphics[width=\linewidth, trim = 0cm 0cm 0cm 0cm]{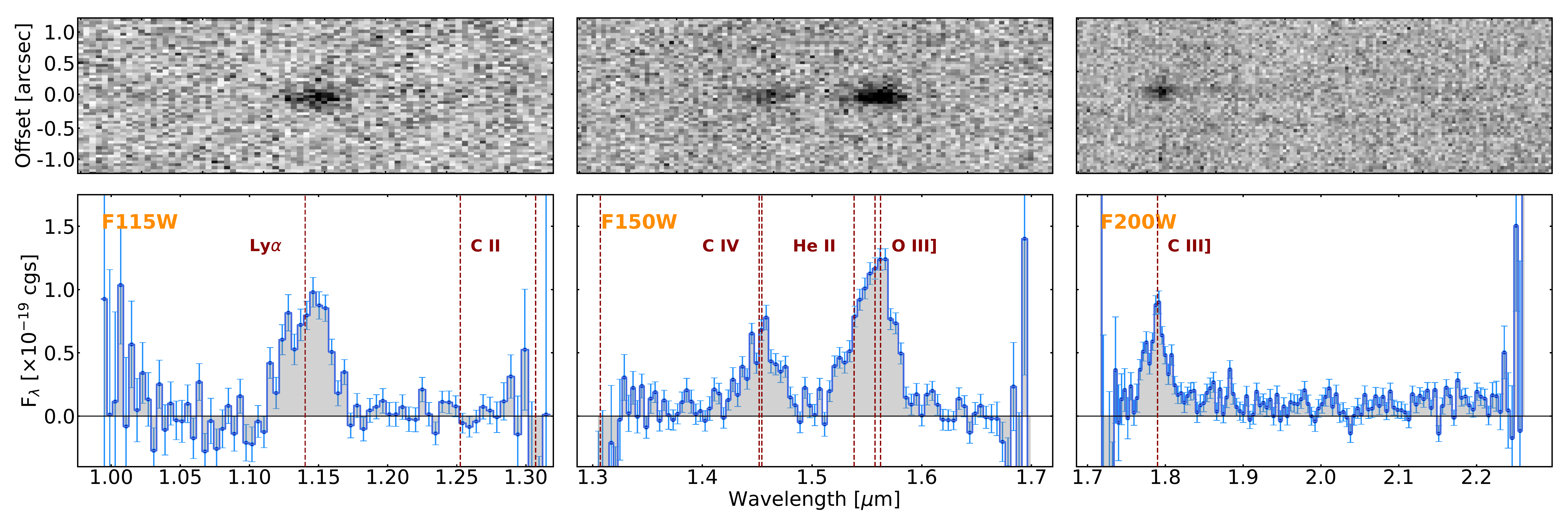}
\caption{A simulated $\sim26.5$ AB mag $z=8.38$ galaxy \citep{laporte17} showing Ly$\alpha$, \CIV, \HeII. \OIII\ and \CIII\ as observed with our NIRISS program and simulated with \mirage\ and \grizli\ (using the full set of exposures from both GR150C and GR150R dispersers). The input SED is taken from the ASTRODEEP catalogs and artificial emission lines are added for illustration: \lya\ flux is assumed to have an integrated flux of $1\times10^{-17}$ \ergscm, while all other rest-frame UV lines have $4\times 10^{-18}$ \ergscm\ with simple Gaussian profiles \label{fig:austin}. From left to right, the panels show results using the F115W, F150W and F200W filters, respectively. No contamination by foreground galaxies or intra cluster light, and no lensing magnification, have been included in this simulation. Simulations including the effects of lensing magnification and foreground contamination are presented in Section~\ref{sec:NIRISS}, and Figures~\ref{fig:niriss_sims} and~\ref{fig:lya_complete}.}
\end{figure*}

The GLASS-JWST-ERS program was designed to shed light on these issues by enabling the following measurements: 

\begin{enumerate}

\item \textit{\lya\ systemic velocity offsets at $z>6$.} Figure~\ref{fig:NIRSPEC_LYA1} highlights the need for a high resolution ($R\sim2700$) grating with NIRSpec in order to obtain sufficient spectral resolution and wavelength coverage with which to directly measure \lya\ velocity offsets and line profiles with respect to the galaxy systemic velocities traced by optical lines. Such data are crucial for determining ISM/IGM absorption and are currently completely unknown except for a handful of sources at these redshifts \citep[e.g.,][]{Sta++17,Pentericci2016ApJ...829L..11P,pentericci18,Endsley2022arXiv220201219E}.  

\item \textit{\lya\ spatial extent.} In combination with the velocity offsets, spatially resolved information is needed to differentiate between \lya\ resonant ISM vs.\ IGM scattering.  This will be provided by the NIRISS observations.

\item \textit{\lya\ and UV continuum spatial offsets.} These will also yield a slit-loss distribution for NIRSpec (Figure~\ref{fig:NIRSPEC_LYA2}), which is unknown but crucial for interpreting $z>6$ \lya\ fluxes \citep{Hoag2019MNRAS.488..706H,Lemaux:2021}.

\item \textit{Rest-frame optical line redshifts, and optical line measurements for LBGs with undetected \lya.} The vast majority of $z>7$ dropouts lack spectroscopic confirmation. Luminosity functions are thus based entirely on the robustness of dropout selection techniques. Furthermore, Spitzer/IRAC data suggest that some $z>7$ galaxies may have unusually strong optical emission lines indicative of especially young and highly star-forming systems \citep{Laporte14,labbe13,Smi++14,rb16,Castellano2017,Debarros2019}, or even mature stellar populations governed by strong Balmer breaks \citep{hashimoto18,tamura19,rb20,strait20,laporte21}. Our NIRSpec data will provide the first wholesale empirical test of these hypotheses. 

\item \textit{UV emission line fluxes.} We will detect or tightly constrain \ion{C}{3}] and \ion{C}{4}. These lines can be enhanced at high-$z$, possibly due to extreme metal-poor stellar populations \citep{Sen++17,Sen++22,Sta++17,Hutchison2019ApJ...879...70H,Berg2019,vanzella2021,Feltre:2020}. In addition to metallicity and star formation conditions, the detection of these lines may aid in identifying high-$z$ active galactic nuclei candidates, using diagnostic diagrams based on \ion{He}{2}, \ion{C}{4}, \ion{C}{3}], and \lya\ \citep{Feltre2016,Nakajima2018,laporte17b}. As an example, we show a simulated $z=8.38$, $\sim26.4$ AB mag galaxy in the cluster field \citep{laporte17} as observed with our NIRISS setup in Figure~\ref{fig:austin}, where each of these lines are visible.

\item Based on current estimates of the luminosity function \citep{MTT15,McLeod:2015,bouwens15,oesch18}, \textit{rest-frame UV/optical photometry and sizes for $\sim$100-200 LBGs at $z\gtrsim7$ will be obtained with NIRCam in parallel mode.} The images will provide new information on their abundances, stellar masses and star formation histories \citep{rb21a}, as well as on the size luminosity relation \citep[][see also Yang et al. 2022, submitted]{Grazian2012,Bowler2017,Bouwens21}. Based on the NIRcam point spread function width at 2$\mu$m of $\sim$80 mas, star-forming clumps can be probed
down to 500-300 pc at $z=6-12$, and down to $<100$ pc with the assist of lensing magnification.

\end{enumerate}

\subsection{Key Science Driver 2: How do baryons cycle through galaxies?}
\label{ssec:KSD2}

Why do some galaxies continue to form stars and others do not? What determines the relative growth of stellar disks and bulges? These questions relate to the baryon cycle; the competition between gravity and star formation/black hole-driven outflows (i.e., feedback), thought to regulate star formation in galaxies. Gas-phase metallicities are one of the key probes of this cycle as they are sensitive to the flow of material from galaxies into the circumgalactic medium \citep[CGM; see][for recent reviews]{Maiolino:2019vq,Kewley:2019kf}. Outflows of enriched gas efficiently distribute metals throughout the ISM and CGM via galactic fountains \citep[e.g.,][]{rb19}. Likewise, pristine gas accretion reduces ISM metallicities and enhances star formation.

The slope of the mass--metallicity relation and its SFR-dependence are sensitive probes of feedback and the baryon cycle \citep{DFO12,Hen++13a,Hen++13b,Henry:2021,Sanders:2021ga}, as is the spatial distribution of metals \citep[i.e., radial gradients;][]{Jon++13,Ang++14}.  Models make robust but conflicting predictions for such gradients and the mass--metallicity relation depending on the strength of feedback and its outflow properties \citep{Gib++13,Ma++17,Tissera:2018vv,Hemler:2021bw}. Tension is highest at low masses ($\log M_*\lesssim8$), especially at $z>2$ \citep{Pil++12,Pil++17,Wang:2016um,Wang:2020bp}.  ``Inside-out'' growth models imply steep gradients that flatten at later times and at higher masses as disks grow. Other scenarios suggest metals should be well-mixed by early feedback, then locked into steep gradients as winds lose the power to disrupt massive gas disks \citep{M+D04,Few++12,Ma++17}.  Stronger feedback should also lead to a steepening mass--metallicity slope at low masses \citep{Hen++13b,Henry:2021ju,Wang:2022MZR}. Thus, high-$z$ observations of the low mass end of the mass-metallicity relation would be particularly valuable in discriminating between competing models.


\begin{figure*}
\includegraphics[width=0.99\linewidth, trim = 0cm 0cm 0cm 0cm]{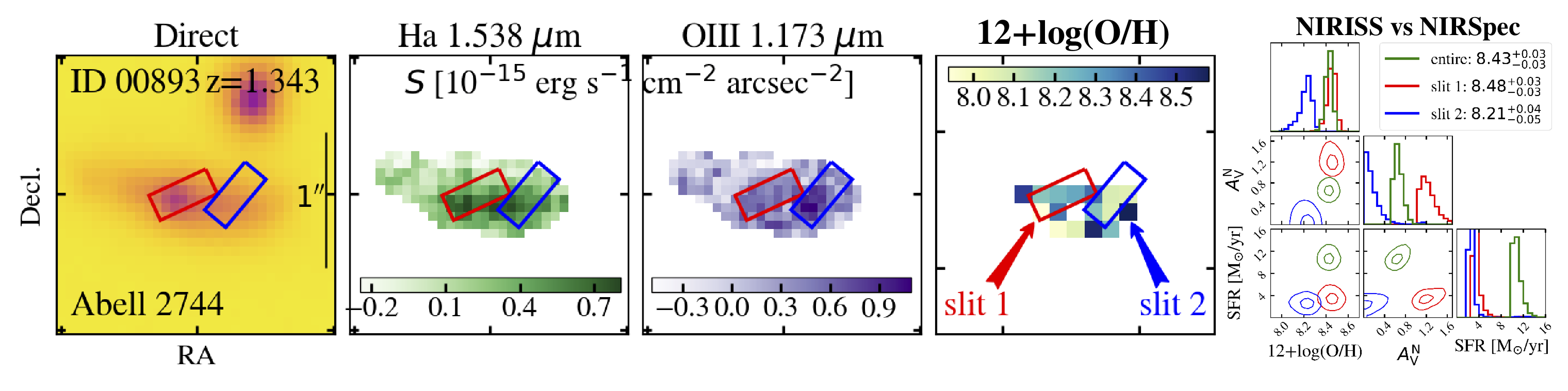}
\caption{A $z=1.34$ galaxy magnified by A2744, illustrates the power of combining NIRISS and NIRSpec data. {\it From left:} F160W image, and GLASS H$\alpha$, [\ion{O}{3}]$\lambda5007$, and gas-phase metallicity maps \citep{Wang:2016um,Wang:2019cf,Wang:2020bp}. Two NIRSpec slits
($0\farcs2\times0\farcs46$) centered on the F160W and [\ion{O}{3}] peaks are shown as red/blue boxes. A forecast of the gas-phase metallicity, nebular dust attenuation, and SFR estimates derived from the proposed full-galaxy NIRISS (green), and partial-galaxy NIRSpec data is shown at {\it right}. Inferences change
dramatically with aperture: metallicity varies as much as 0.4 dex (2.5$\times$) between the two slits. Our sample will showcase the variations within sources and help build statistical recipes to account for aperture effects.
\label{fig:xin}}
\end{figure*}

\smallskip
Currently the $z>4$ mass-metallicity relation is essentially unknown \citep{Jon++20,Sha++17}, and most $z>1$ spectroscopy is spatially unresolved. Recently, progress has been made in obtaining a large homogeneous sample of $\lesssim$kpc-resolution metallicity gradients at cosmic noon using \hst\ WFC3/IR grism spectroscopy \citep{Wang:2016um,Wang:2020bp,Simons:2020wp,2022arXiv220403008L} or ground-based integral-field spectroscopy supported by adaptive optics \citep{Leethochawalit2016,ForsterSchreiber:2018uq}. However, these efforts are exclusively focused on the $z\lesssim2.5$ universe, given the limitation of current instrumentation.

To distinguish what kind of feedback is active in which galaxies and when, we must map metallicity, dust, and SFR for systems spanning large mass and redshift ranges ($\log{M_*}\sim6$--10 and $z\gtrsim2$, when disks/bulges emerged and feedback was most active). Deep, spatially resolved spectra of large samples is the only route to this understanding. \jwst\ uniquely provides sensitive, uninterrupted wavelength coverage and thus the necessary sets of multiple diagnostic lines required for this
analysis.


GLASS-JWST-ERS was designed to shed light on these issues by enabling the following measurements. 

\begin{enumerate}
\item With NIRISS, \textit{ionized gas metallicity, dust extinction, and SFR maps will be obtained in $\sim$50 $z\lesssim3.5$, $\log{M_*}\gtrsim6$ galaxies}, sampling relatively low masses where current models diverge most. These maps will be crucial for interpreting NIRSpec data, which necessarily sample only subsets of galaxies. The wavelength coverage will enable detection of the optical lines between [\ion{O}{2}] and H$\alpha$ at $1.7<z<2.3$, of the lines between [\ion{O}{2}] and [\ion{O}{3}] at $1.7<z<3.4$.
\item NIRSpec will \textit{spectrally resolve key diagnostic lines ([\ion{N}{2}] + H$\alpha$, the auroral [\ion{O}{3}] line at 4363\AA\ and H$\gamma$, [\ion{Ne}{3}] + \ion{He}{1} + Balmer lines, and doublets such as [\ion{S}{2}] and [\ion{O}{2}])} which are blended at the lower resolution of grism spectra. This addresses the major limitation of grism-only data by improving dust attenuation estimates and detecting weak AGN which might otherwise bias results.
\item NIRSpec will \textit{measure metallicity, dust attenuation, and SFR in $z\gtrsim4$ galaxies}, probing the baryon cycle at these epochs for the first time \citep{Henry:2021ju,Wang:2022MZR}.
\item The combination of NIRSpec and NIRISS will \textit{cover multiple rest-frame optical metallicity diagnostics for each galaxy}. While metallicity calibrations remain uncertain at high redshifts \citep{Kew++13}, these data will enable comparisons between different diagnostics and support detailed photoionization modeling \citep{Ste++16,Che++18}. An additional goal is to detect or set stringent upper limits on the auroral [\ion{O}{3}] line at 4363\AA\ in the galaxies with the brighter lines, providing  temperature-based metallicities in these cases.
\end{enumerate}

\subsection{Ancillary science cases}
\label{ssec:ancillaryscience}

The high spatial and spectral resolution public data our program will deliver will support a broad range of external investigations. Here we detail a few examples, which will hopefully represent just a small subset of the community driven investigations:

\begin{enumerate}
    \item NIRISS will provide Pa$\beta$, Pa$\gamma$ and [\ion{S}{3}] maps for all of the cluster member galaxies selected from the Grism Lens Amplified Survey from Space \citep[GLASS][]{Tre++15,Vul++16,Vul++17}, and H$\alpha$---and thus SFR--- Pa$\beta$ and [SIII] maps for field galaxies at $z>0.5$, enabling the extension of previous resolved  SFR, dust and metallicity studies to incorporate this critical piece of information pertaining to feedback processes. 
\item NIRSpec will provide Balmer decrements which---in combination with extant rest-UV photometry from HFF (and redder GTO NIRCam data)---will shed light on the dust content and distribution of high-$z$ galaxies. This is a critical systematic in stellar mass and SFR estimation.
\item NIRSpec will provide Pa$\alpha$ and other IR lines (e.g. He I, CaII) for cluster galaxies which will shed light on the properties of the central photoionization source, allowing to study in detail SFR and AGN. 
\item NIRISS + NIRSpec will support extensive refinements of photometric redshift estimation techniques by providing a robust spectroscopic redshift and template training database at higher redshifts than are currently available.
\item NIRISS will enable unprecedented spectroscopic continuum studies, including measuring
Balmer and metal absorption features (e.g., Mg$b$) at $z>2$, and perhaps allowing their radial gradients to be inferred.

\item The new redshifts and 2D spectra will improve lens models and help investigate the nature and distribution of dark matter.
\item The comparison of our NIRISS data with previous \hst\ images will provide another epoch for the detection of unusual transients, such as multiply imaged supernovae \citep{Kelly2015,Treu2016} or highly magnified stars \citep{Kelly2018}.
\end{enumerate}

\section{Target Selection}
\label{sec:target}

\subsection{Field Selection}

As demonstrated by multiple programs, including the HFF campaign, observing a lensing cluster has multiple advantages over a blank field for understanding the high redshift universe. Background sources are magnified and thus can be studied at higher depth and higher intrinsic resolution than in a blank field. Even at the median magnification of $\mu=2.3$, the gain in exposure time needed to reach the fainter galaxies, that are the most sensitive probes of reionization and of the baryonic cycle, is substantial (the required exposure time needed to match the depth without magnification would be longer by a factor $\mu^2$). In the regions with highest magnification, lensing allows one to probe one or two order of magnitude fainter sources and to achieve angular resolution as high as 10-20~pc \citep{Van++17,Bouwens2017,Cava2018}. Furthermore, pointing at a cluster enables cluster science and dark matter science via lensing, within the same set of data used for background sources.

A2744 was selected as the GLASS-JWST-ERS primary target for the following reasons: {\it i)} As a HFF cluster, it has exquisite ancillary ultra-deep \hst\ imaging \citep{Lotz2017}, GLASS \hst\ \citep{Tre++15} and ground-based spectroscopy \citep{Mah++17}; {\it ii)} Its publicly available lens models at the time of selection (2017) were some of the most robust ever produced, being based on 83 spectroscopic multiple images \citep{Mah++17, ric+21}. A new detailed strong lensing model will soon become available (Bergamini et al., in prep.). The new model reproduces with a significantly lower rms offset ($\approx 0.4 \arcsec$) the observed positions of 90 spectroscopically confirmed multiple images, with secure optical counterparts from \hst\ or VLT/MUSE, from 30 different sources with redshifts between 1.69 and 5.73. In this new study, the cluster total mass distribution is modeled by exploiting the measured stellar velocity dispersions of the brightest cluster members 
(85 among the 225 selected member galaxies, of which 202 are spectroscopic)
and informative priors on the mass of three external substructures detected through an independent weak lensing analysis \citep{Medezinski2016}; {\it iii)} It is visible during the ERS window; {\it iv)} It is slated for GTO imaging, which will aid spectroscopic interpretation; {\it v)} It has minimal galactic extinction (A$_{\rm F125W}=0.0118$), implying low infrared background.

\subsection{NIRSpec target selection}\label{sec:msatarg}

\begin{figure}
\includegraphics[width=\columnwidth, trim = 0cm 0cm 0cm 0cm]{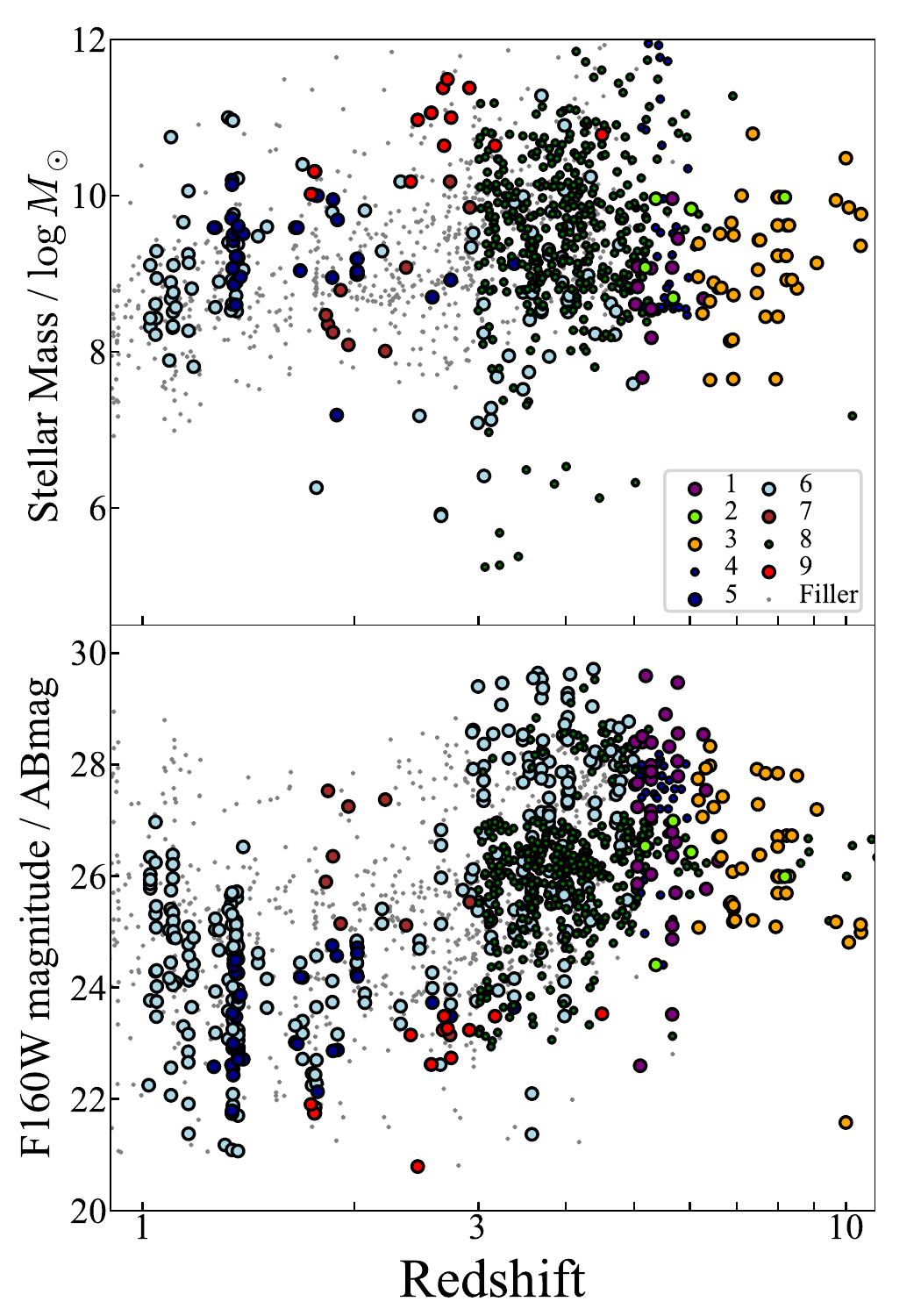}
\caption{
    Redshift distributions of the primary science target galaxies (filled circles; Sec.~\ref{sec:msatarg}) and fillers (gray) for the NIRSpec Micro-shutter assembly observations. The colors identify the samples as described in Section~\ref{sec:msatarg}. With NIRISS we will take spectra of most of them.
    }
\label{fig:summary}
\end{figure}

Based on spectrophotometric catalogs from our team and the literature, the following groups of targets (ordered by decreasing priority) were identified at the time of submission. Target selection was updated and refined over the years to take advantage of new data, both spectroscopic \citep{ric+21} and imaging follow-up campaigns \citep{stein20}. Sources collected from literature are, when necessary, cross-matched to our \hst\ catalog, which is aligned to the Gaia-DR2 wcs frame, to refine the coordinates. Figure~\ref{fig:summary} shows the redshift, $m_{\rm F160W}$, and stellar mass distributions of the primary samples, based on catalogs available at the time of this writing (early 2022). NIRSpec micro-shutter assembly \citep[MSA][]{NIRSPEC2} configurations will be designed around them, according to the constraints provided by the schedule. We anticipate of order 50 galaxy spectra will be included in the MSA configuration. In addition, spectra of relatively bright galaxies and stars will be included as secondary calibrators. Our primary science targets include the following populations. The design of the MSA configuration will aim to obtain spectra representing these classes, with emphasis on those relevant for the key science drivers.

\begin{enumerate}
\item $z > 5$ spectroscopically-confirmed galaxies (sample size 47).
\item $z > 5$ extremely high {\it Spitzer}/IRAC-inferred EW(H$\alpha$) -and EW(\OIII) galaxies (5).
\item  $z > 6$ photometrically-selected galaxies, including a $z\sim8$ proto-cluster (48).
\item $5 < z < 6$ photometrically-selected galaxies (46).
\item  $1 < z < 2$ spectroscopically-selected emission line galaxies from GLASS  (30).
\item $1 < z < 5$ spectroscopically-confirmed galaxies (327).
\item $1.1 < z < 3.4$ photometrically-selected galaxies (9).
\item $z > 3$ photometrically-selected galaxies outside the HFF WFC3 footprint (411).
\item $1.7 < z < 4$ expected continuum sources from HFF imaging and GLASS spectroscopy (13).

\end{enumerate}

In addition, we include photometrically-selected sources, as well as those located outside of the WFC3-IR coverage but within the ACS footprint, as fillers. We will also sample the intracluster light and if possible obtain spectra of some cluster members.

\section{NIRISS observations}
\label{sec:NIRISS}

In this section we first describe the NIRISS observations (\S~\ref{ssec:niriss_setup}) and then present a suite of simulated data sets that we use to forecast the expected sensitivity (\S~\ref{ssec:niriss_sims}). We anticipate that the simulations will not match exactly the flight performance, but they should provide a sufficient approximation for the purpose of planning work on the public data sets. The NIRISS setup is summarized in Table~\ref{tab:primary}.

\begin{table}
    \centering
    \begin{tabular}{llllcc}
    \hline
    Instrument     & Mode    & Filter & Disp.   & Exptime & Limit \\
                   &         &        &         &   (s)   & 5-$\sigma$ \\
                   \hline
    NIRISS         & Imaging & F115W  &  -      & 2830    & 28.6 \\
    NIRISS         & Imaging & F150W  &  -      & 2830    & 28.7 \\
    NIRISS         & Imaging & F200W  &  -      & 2830    & 28.9 \\
    NIRISS         & WFSS    & F115W  & GR150R  & 5200    & -\\
    NIRISS         & WFSS    & F115W  & GR150C  & 5200    & -\\
    NIRISS         & WFSS    & F150W  & GR150R  & 5200    & -\\
    NIRISS         & WFSS    & F150W  & GR150C  & 5200    & -\\
    NIRISS         & WFSS    & F200W  & GR150R  & 5200    & -\\
    NIRISS         & WFSS    & F200W  & GR150C  & 5200    & -\\
\hline
    NIRSpec        & MSA     & F100LP & G140H  & 17682    & -\\
    NIRSpec        & MSA     & F170LP & G235H  & 17682    & -\\
    NIRSpec        & MSA     & F290LP & G395H  & 17682    & -\\
    \hline
    \end{tabular}
    \caption{Summary of primary observations to be carried out in the field of A2744. For the NIRISS acquisition images we provide the estimated depth based on the exposure time calculator. The sensitivity of the spectroscopic observations is discussed in the text, Sections~\ref{sec:NIRISS} and~\ref{sec:NIRSPEC}.}
    \label{tab:primary}
    \end{table}

\subsection{NIRISS observational setup}
\label{ssec:niriss_setup}

The two orthogonal $R=150$ grisms in the F115W, F150W, and F200W filters will provide continuous wavelength coverage in the wavelength range 1--2.2 $\mu$m. The spectra will overlap with those obtained by NIRSpec at 1--2.2$\mu$m, enabling direct, quantitative comparisons of spatially and spectrally resolved spectroscopy in that range.

NIRISS provides two orthogonal spectra, replicating the strategy adopted by GLASS \citep{Sch++14,Tre++15} to mitigate contamination by nearby objects. This is especially important in the cluster core. With GLASS, 64\% of the objects had at least one position angle that was free of contaminants \citep{Tre++15}. Since NIRISS spectra are shorter, we expect smaller levels of contamination: indeed, we find 84-91\% of all simulated spectra (see details of their construction below) have ``mild'' contamination, defined as such in the case of $<$10\% of detector pixels containing contaminating (neighbouring) galaxy flux at the $5\sigma$ level. The top end of this range represents values found for the F115W filter, while the lowest estimates represent levels found for the F200W, which covers by far the largest wavelength range (almost a factor of 2$\times$ larger than the F115W filter) and thus detector area. If we consider spectra with such mild contamination as ``clean'', then harboring the power of the orthogonal setup afforded by the two grisms adopted by the ERS program, we find that $\sim$90-97\% of sources have at least one clean spectrum. The contamination level will of course increase for fainter objects, but these will largely be emission-line-only sources, for which contamination and deblending models are robust (as demonstrated by GLASS and NIRISS simulations we have carried out using the Grism Redshift \& Line\footnote{\grizli: \url{https://grizli.readthedocs.io/en/latest/}} analysis software for space-based slitless spectroscopy).

The total exposure time per filter per grism is divided into 4 exposures of 1300s each with small dithers between them, using both GR150R and GR150C grisms, for a total of 2.9 hours per filter.  As part of the Wide Field Slitless Spectroscopy (WFSS) observing sequence we obtain four direct images per filter (before and after each orthogonal grism with extra dithers to cover the FOV of both grisms) of 350s each for a total observing time of 2830s per filter, used to determine the flux, trace position and wavelength in the grism data. The expected sensitivity of the images based on pre-flight ETC is given in Table~\ref{tab:primary}.

\subsection{Simulated data sets}
\label{ssec:niriss_sims}

To construct accurate simulations of our observational program, we utilize a combination of \mirage\ (see Appendix~\ref{sec:tools}) and \grizli. The construction of the simulated data sets is divided into three main stages, namely the simulation of (i) accurate science and dark frames (both direct and dispersed), (ii) the post-processing through the \jwst\ data reduction pipeline, and (iii) the injection of lensed, high-$z$ background sources for estimations of emission line completeness.

\subsubsection{Generation of NIRSS simulations}

For the first step and construction of NIRISS science images, we adopt the \hst\ $H_{\rm 160}$-band observations from the ASTRODEEP\footnote{\url{http://www.astrodeep.eu/}} Frontier Fields catalogs of A2744, which offer intra-cluster light (ICL) -subtracted images and photometry of the cluster and background galaxies, as well as their modelled SEDs. Additionally, the catalogs also offer a modelled postage stamp of the ICL with an associated model SED (which we assume to be representative for the entire ICL). We thus use the $H_{\rm 160}$-band (background galaxies, lensing cluster and the ICL) and rescale each of the postage stamps to the required countrate as defined by their apparent magnitude in either the F115W, F150W or F200W NIRISS filter (as measured through the convolution of each galaxy's SED and the filter response curve) and the relevant PHOTFNU value. All non-galaxy pixels are then set to zero to simulate a noiseless NIRISS image based on galaxy SEDs, and the resulting countrate images of each galaxy summed together to create the final noiseless scene. The final countrate images, segmentation maps and galaxy SEDs are then passed to \grizli\ which re-orients a given scene to the \jwst\ pointing of choice (as defined by an input \texttt{yaml} observation file) and disperses the spectra according to the NIRISS grating (GR150C or GR150R) associated with the exposure. The resulting (noiseless) direct and dispersed images from \grizli\ are then passed back to \mirage, which combines the noiseless images with realistic dark current (based on the observational setup and exposure time of the ERS program) to create the final, uncalibrated exposure in the manner described in the Appendix. We note that in our simulations the effects of optical ghost sources\footnote{\url{https://jwst-docs.stsci.edu/jwst-near-infrared-imager-and-slitless-spectrograph/niriss-predicted-performance/niriss-ghosts}} are not included.

\subsubsection{Data processing}

\begin{figure*}
\center
 \includegraphics[width=\textwidth]{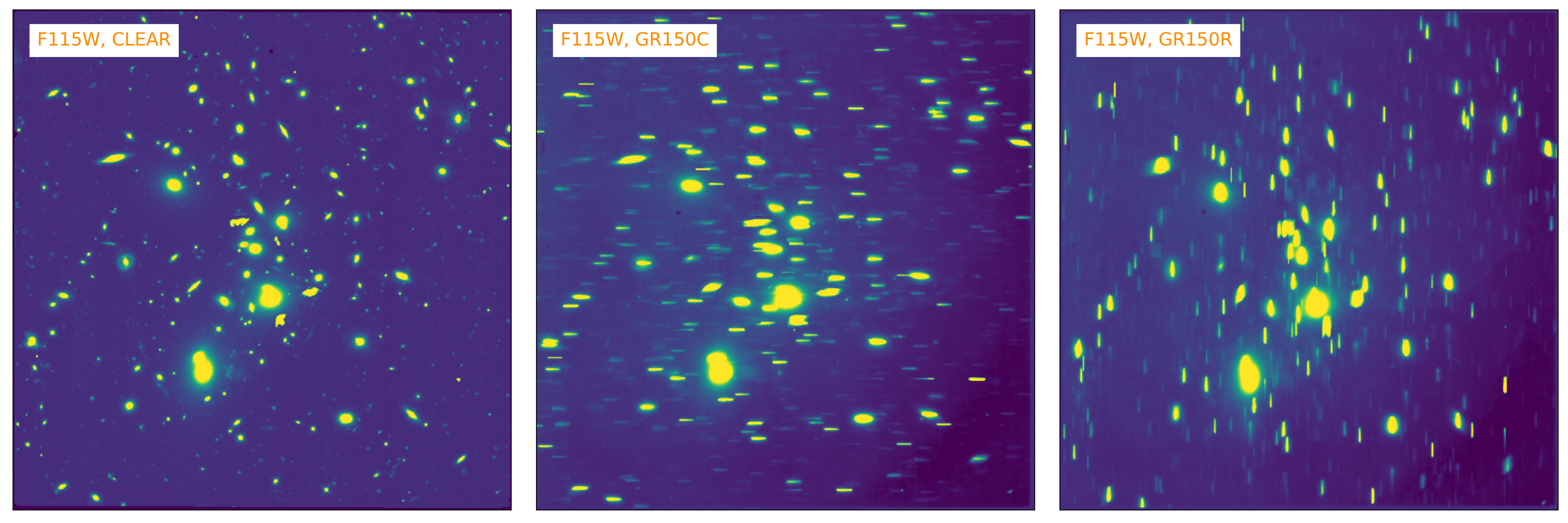}
 \caption{Simulated F115W NIRISS images (direct imaging, GR150C and GR150R configurations, respectively) of the Abell 2744 cluster produced using a combination of \mirage\ and \grizli\ and reduced using the public STScI data reduction pipelines.}
 \label{fig:niriss_sims}
\end{figure*}

As part of the second step to reduce the resulting uncalibrated exposures, we use the latest version (v1.4.3) of the \jwst\ data reduction pipeline in identical fashion to the steps described in Appendix~\ref{sec:tools}. Briefly, direct images are run through all of the available image pipelines to create a drizzled and astrometrically- and photometrically-calibrated image for each filter, with an output catalog of detected sources. That catalog is subsequently used to locate the position of sources in the spectroscopic exposures in order to correctly estimate and subtract contaminating background from the spectra. To obtain a final, co-added spectroscopic image, we align each of the WFSS exposures and take the median of the 2D images to obtain our final image with each filter/grism combination. Examples of the final images we produce (direct and dispersed using the F115W filter) are shown in Figure \ref{fig:niriss_sims}.

\subsubsection{Line flux completeness simulation}

Finally, for completeness estimates of emission lines we also simulate the recovery of mock luminous, high-$z$ ($z\geqslant$8) sources. We use \lya\ as a a scientifically compelling example, but we stress that the sensitivity/completeness estimates presented here apply to any spectrally unresolved and spatially compact lines of the same flux and wavelength.

We begin by creating mock catalogs of background galaxies based on a grid of redshifts and Ly$\alpha$ emission. We create simple artificial SEDs of the high-$z$ galaxies based on a simple line profile over a redshift grid of $z=8-15$ in integer intervals (except for $z=13$, where Ly$\alpha$ falls between the F150W and F200W bands) and an integrated flux grid of f(Ly$\alpha$)=$1-5000\times10^{-18}$ \ergscm\  (specifically f(Ly$\alpha$/1$\times10^{-18}$) values of 5, 10, 50, 100, 500, 1000 and 5000 \ergscm). For simplicity, we assume Gaussian profiles for the line emission which is placed at the redshifted Ly$\alpha$ wavelength. To keep our results applicable to virtually any emission line of interest, we neglect IGM absorption effects which would typically affect the measured Ly$\alpha$ line flux.
Source-plane images of the galaxies are then created using postage stamps, where we assume circular profiles with an on-sky radius of $\sim0.16''$ (corresponding to a median physical size of 0.8 kpc for luminous galaxies at $z\sim8$, as measured in the latest results from the Brightest of Reionization Galaxies Survey; \citealt{rb21b}) and a pixel resolution of 0$\farcs$05 arcsec/pixel, with a FOV of 150$\times$150 arcsec$^{2}$ and 100 mock galaxies within the area. 

Given their positions behind the lensing cluster, we aim to simulate lensing effects on the high-$z$ galaxies by the low-redshift galaxy cluster through the use of the gravitational lensing code \texttt{Lenstronomy} and the (NFW) deflection maps of \citet{Zit++15}. We begin by scaling the deflection maps by the lensing efficiency ratio at the desired redshift of the background galaxies (i.e., by the ratio of the angular diameter distance between the cluster and background galaxies, and the angular diameter distance to the background galaxies). Each galaxy is then lensed via a backward ray-tracing method according to the deflection maps, from the source plane to the image plane pixel grid given by the ASTRODEEP NIRISS image defined above. The lensed galaxy images and associated emission line spectra are then passed to \grizli\ for dispersion and orientation to each ERS pointing, in identical fashion to the ASTRODEEP images, and subsequently added to the reduced image products after being converted to the appropriate units. The above procedure is repeated 100 times for each redshift/f(Ly$\alpha$) combination. 

To estimate the resulting completeness of our simulations, we define completeness as the median fraction (over the 100 iterations) of recovered sources in the direct images and the number of those recovered sources with detected Ly$\alpha$. For the latter consideration, we cross-match the direct imaging catalog (derived at the \texttt{calwebb\_image3} stage) with the dispersed spectra and integrate the background-subtracted spectrum over the expected position of Ly$\alpha$ while estimating the noise as the standard deviation of the sky adjacent to the source multiplied by the square-root of the number of pixels considered. This method allows us to account, in part, for contamination of the high-$z$ source by neighbouring foreground sources if the former's flux were to be significantly contaminated by the latter. 

\begin{figure}
\center
 \includegraphics[width=\columnwidth]{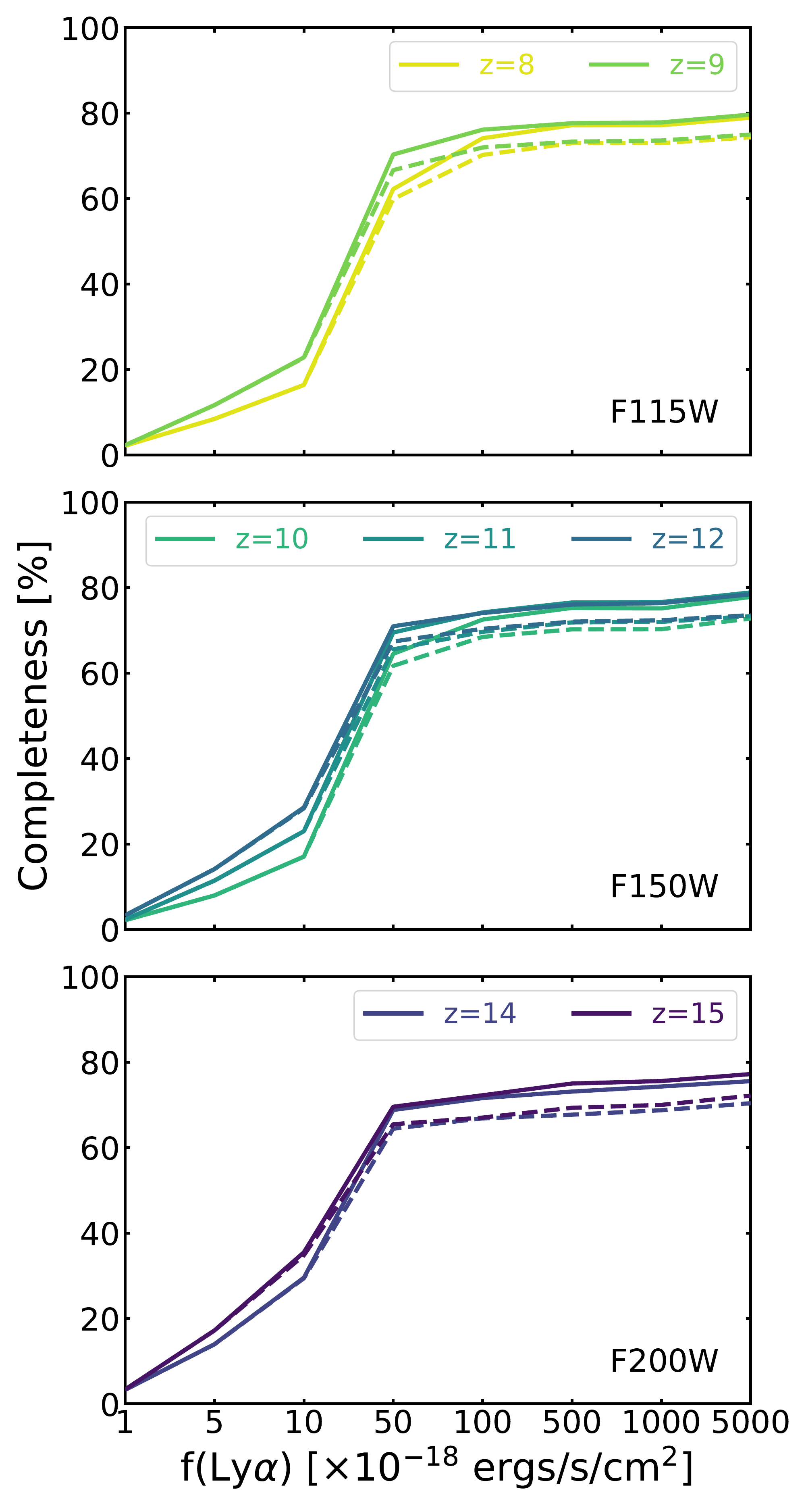}
 \caption{Estimates of galaxy and Ly$\alpha$ emission line completeness with NIRISS direct imaging and wide field slitless spectroscopy, respectively, based on the recovery of source and line profiles for mock sources and SEDs into realistic images. Solid lines denote the fraction of recovered sources in direct imaging with the F115W (top), F150W (middle) and F200W (bottom) filters, while the dashed lines represent the fraction of those detected sources with detected Ly$\alpha$ emission. At the bright end completeness saturates at $\sim80$\% due to the effects of contamination by foreground cluster galaxies and intracluster light. At the faint end, completness declines gently due to the effects of gravitational lens magnification that allows us to detect lines intrinsically fainter than in blank fields. The results are quoted for \lya\ but they apply to any other emission lines in the background of the cluster, at the corresponding wavelength, except for small differences due to dependency of lensing magnification on redshift. }
 \label{fig:lya_complete}
\end{figure}

The procedure is performed for both grism setups and we consider a source to be detected in Ly$\alpha$ if  the extracted line profile has an integrated signal-to-noise ratio $>5\sigma$ in only one of the two configurations or $>3\sigma$ in both, and display the results of our simulations in Figure~\ref{fig:lya_complete}. We find our imaging completeness remains largely constant at the $\sim70-80$\% level, primarily due to the apparent brightness of the sources which reach magnitudes of $\sim21-27$ AB with the strongest \lya\ ($>5\times10^{-17}$ \ergscm). We find that even for very bright emission lines the completeness saturates at 80\%: 20\% of the lines are lost due to contamination by foreground cluster galaxies. By extension, we find our Ly$\alpha$ completeness estimates decrease towards fainter line fluxes as expected, dropping down to $\sim$40\% for f(Ly$\alpha$)=$1\times10^{-17}$ \ergscm\ and down to $\sim2-5$\% for even fainter line fluxes.  We note that, owing to the effect of lensing magnification, we our completeness is non zero down to fluxes of order 10$^{-18}$ \ergscm.

While these simulations represent an idealized scenario in terms of the treatment and data reduction of direct and dispersed images from which sensitivity estimates may be derived, they also naturally fold in the effects of flux contamination and confusion by the massive cluster galaxies and ICL, thus representing realistic simulations of expected data sets. Additionally, while this exercise has been performed for high-redshift galaxies and Ly$\alpha$ emission to demonstrate a full suite of simulations accounting for ICL, lensing and data reduction steps, the completeness results are sufficiently generic to be applicable to other emission lines at a variety of other redshifts.

\section{NIRSpec Observations}
\label{sec:NIRSPEC}

In this section we first describe the NIRSpec observational setup (\S~\ref{ssec:nirspec_setup}) and then present some examples of spectra expected from this program (\S~\ref{ssec:nirspec_sims}). We anticipate that these spectra will not match exactly the flight performance, but they should provide a sufficient approximation for the purpose of planning work on the public data sets. The NIRspec observations are summarized in Table~\ref{tab:primary}.

\subsection{Observational Setup}
\label{ssec:nirspec_setup}

GLASS-JWST-ERS will use the $R=2700$ gratings G140H, G235H, G395H, sufficient to resolve \lya\ and measure systemic velocities.  At this resolution, we expect most other nebular lines will be marginally resolved or unresolved \citep{Mas++14}.  Hence, in these cases high-resolution maximizes our sensitivity.  Spectral coverage from 1-5 \micron\ will enable the detection of features between [\ion{O}{2}] and H$\alpha$ at $1.7<z<6.5$, of the range between near UV rest frame and H$\alpha$ at $5 < z< 6.5$, and of the range between Ly$\alpha$ and [\ion{O}{3}] at $7.3 < z < 9$.

Based on our key science drivers we chose to expose for 5 hours in each of the three high-resolution gratings: G140H/F100LP, G235H/F170LP, and G395H/F290LP. 


For dithering, we choose to nod in the 3-shutter slitlet.  We expect that the standard pipeline subtraction of each nod will be inappropriate for the extended galaxies that we are observing. However, this approach is an efficient way to obtain in-slitlet dithering with lower overheads and higher multiplexing than would be possible with standard dithering via additional MSA configurations.  
 At the time of flight-ready program submission, we will include extra background shutters in empty parts of the MSA, in order to ensure reliable background subtraction. These shutters will then be specified as the background in a customized re-processing of the NIRSpec observations.  We will also obtain an additional small sub-shutter dither (``2-POINT-WITH-NIRCAM-SIZE2'') in each exposure, in order to improve PSF sampling of our NIRCam parallel imaging.

We set up our NIRSpec exposures using NRSIRS2, which is recommended for deep observations.  Each NIRSpec band is given 20 groups, 1 integration, and 2 exposures.  This gives six exposure slots, for which we plan NIRCam imaging parallels.  
For NIRSpec, each dither/nod point is then 1473.5 seconds; with 3 nods, 2 sub-shutter dithers, and 2 separate exposures, the total time per band is 1473.5 seconds times 12, or 4.9 hours. 


\subsection{Simulated data sets}
\label{ssec:nirspec_sims}

To simulate results expected from our NIRSpec exposures, we utilize the Python \texttt{Pandeia} exposure time calculation (ETC) engine, upon which the public \jwst\ ETC is constructed. We provide the code with the precise setup designed for our exposures - i.e., a NRSIRS2 readout pattern with observations divided into 20 groups, 1 integration and 2 exposures per band, resulting in a total exposure time of $\sim$4.91 hours per configuration. To provide the most realistic forecasts, we display forecasts of known and spectroscopically-confirmed sources in the cluster. To best illustrate the gain in information and variety of emission lines detected from each configuration, we highlight four real sources at redshifts of $z_{\rm spec}=8.38$ \citep{laporte17}, $z_{\rm spec}=5.054$ \citep{mahler18,ric+21}, $z_{\rm spec}=2.327$ \citep{mahler18,ric+21} and $z_{\rm spec}=1.367$ \citep{Wang:2020bp}.

\begin{figure*}
\center
 \includegraphics[width=0.8\linewidth]{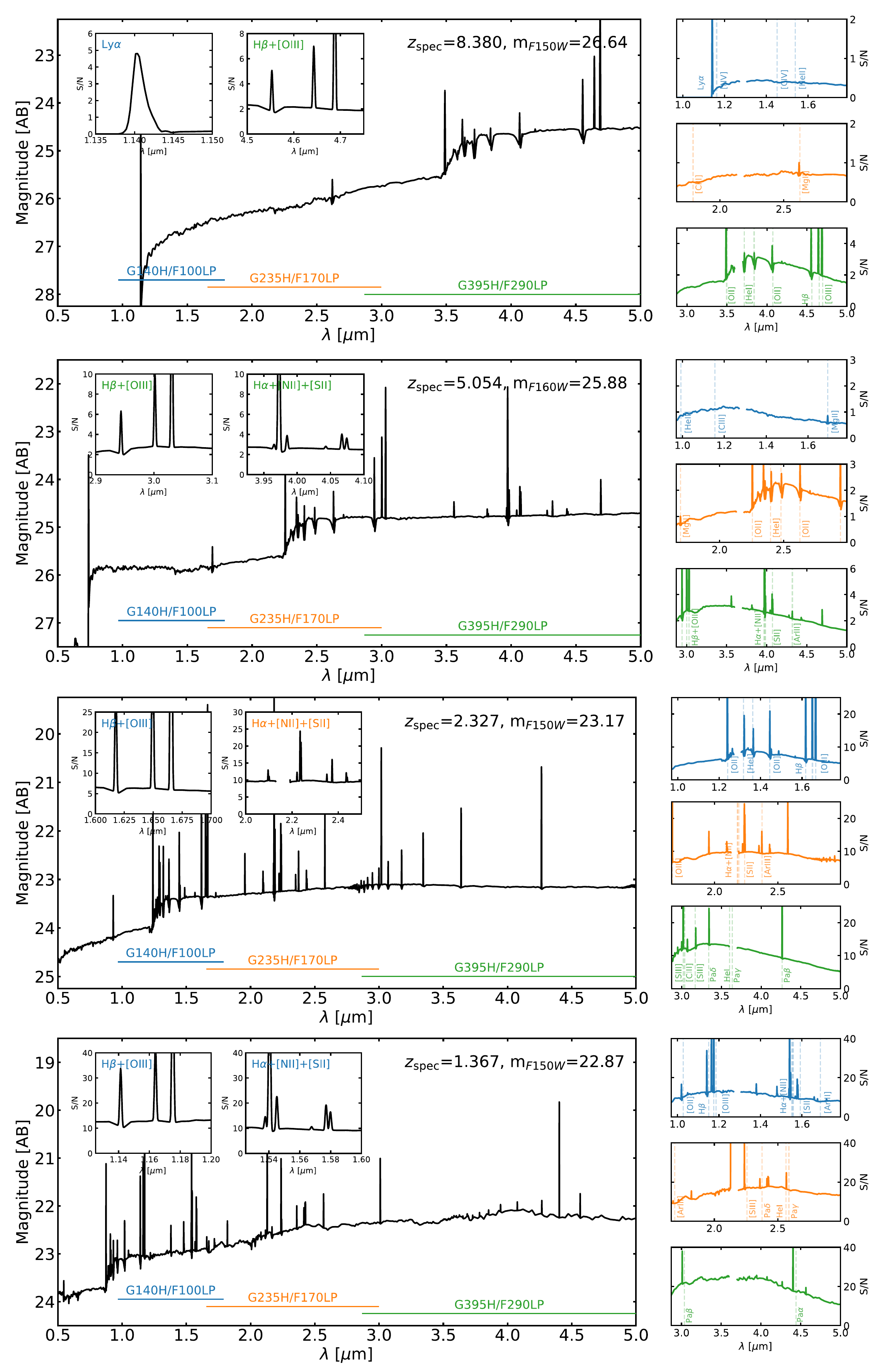}
 \caption{NIRSpec simulations of four spectroscopically-confirmed galaxies within the cluster pointing, based on SED fits to extracted \hst\, VLT and \spitzer/IRAC photometry. Each plot shows a different target galaxy, with the main panel showing the input (noiseless) best-fit SED and each sub-panel highlighting the resulting S/N ratio over the wavelength range probed by a particular configuration (G140H/F100LP in blue, G235H/F170LP in orange, G395H/F290LP in green) and with our observational setup. The two inset panels represent zoom-ins over emission lines of particular interest. The spectroscopic redshift and estimated NIRCam F150W magnitude of each sources are quoted on the top right of the main panels.}
 \label{fig:nirspec_sims}
\end{figure*}

For each source and simulation, we use the high-resolution SEDs from the ASTRODEEP catalogs, while also adding a stacked and skewed Ly$\alpha$ profile from \citet{pentericci18} (normalized to the integrated flux reported by \citealt{laporte17}) to the $z=8.38$ object SED in order to illustrate the resolving power of our setups in determining line profiles. The resulting SEDs are subsequently passed to \texttt{Pandeia} for dispersion according to our \jwst\ observations, assuming the sources are centered in their MSA shutter. We show the results for each of the four targets in Figure~\ref{fig:nirspec_sims}, where we plot the best-fit SED, as well as the (color-coded) NIRSpec spectra. Additionally, for each target we also highlight zoomed-in portions of the resulting simulations to highlight resolved profiles of key emission lines or continuum features. We find for bright galaxies ($m_{\rm F150W}\lesssim26.5$ AB) emission lines such as \OII, \OIII, H$\beta$, and H$\alpha$\ are detected at $>3\sigma$, while the high-resolution spectroscopy delivered by our configurations ensures the lines are also resolved and line profiles of e.g., Ly$\alpha$ can be characterized.

\section{NIRCam parallel observations}
\label{sec:NIRCAM}

In parallel with the spectrosopic observations, GLASS-JWST-ERS will conduct NIRCam imaging covering $\sim18$ sq.\ arcmin in two regions. These data will survey galaxies 3$\arcmin$--8$\arcmin$ (1--2.5 Mpc) from the core of the primary lensing cluster target. This data set will consist of the deepest images taken during the ERS campaign.

We  first describe here the NIRCam observational setup (\S~\ref{ssec:nircam_setup}) and then briefly report on early simulations  to forecast the expected sensitivity (\S~\ref{ssec:nircam_sims}) and prepare the analysis tools. We anticipate that the simulations will not match exactly the flight performance, but they should provide a sufficient approximation for the purpose of planning work on the public data sets. A summary of the NIRCam observations to be obtained in parallel, and the estimated depth based on the pre-flight exposure time calculator, is given in Table~\ref{tab:parallels}.

\begin{table}[]
    \centering
    \begin{tabular}{llll}
    \hline
Primary & Filter & Exptime & Limit      \\
         &        &  (s)    & 5-$\sigma$ \\
   \hline
   NIRSpec & F090W  & 16492   & 29.2 \\
   NIRSpec & F115W  & 16492   & 29.4 \\
   NIRSpec & F150W  & 8246    & 29.2 \\
   NIRSpec & F200W  & 8246    & 29.4 \\
   NIRSpec & F277W  & 8246    & 29.5 \\
   NIRSpec & F356W  & 8246    & 29.6 \\
   NIRSpec & F444W  & 32983   & 29.7 \\
\hline
NIRISS & F090W  & 11520   & 29.0 \\
NIRISS & F115W  & 11520   & 29.2 \\
NIRISS & F150W  & 6120    & 29.1 \\
NIRISS & F200W  & 5400    & 29.2 \\
NIRISS & F277W  & 5400    & 29.3 \\
NIRISS & F356W  & 6120    & 29.4 \\
NIRISS & F444W  & 23400   & 29.5 \\
\hline         
    \end{tabular}
    \caption{Summary of the NIRCam parallels. The coordinates of the field centers are: (0:14:05.5451, -30:20:25:0.28) and (0:13:58.3302, -30:17:58.67) for the two NIRSpec parallels, and (0:14:02:4660, -30:21:37.226) and (0:13:58.3268, -30:18:53.229) for the NIRISS parallels. The NIRISS parallel coordinates are final, the NIRSpec parallel coordinates are not final and they will likely change by a few arcseconds, depending on the final MSA configuration. The expected limiting magnitude is given for a point source based on the exposure time calculator (v.1.7) assuming $0.3-0.5\arcsec$ annulus for background subtraction.}
    \label{tab:parallels} 
\end{table}

\subsection{Observational setup}
\label{ssec:nircam_setup}

GLASS-JWST-ERS takes advantage of NIRCam's capabilities, by simultaneously using the three wide filters in NIRCam's LW arm (F277W, F356W, and F444W) and four wide filters in the SW arm (F090W, F115W, F150W, 200W). The same filter set and a similar balance between the seven filters will be adopted in the NIRSpec and NIRISS parallels, respectively.

The observational setup for the NIRCam imaging taken in parallel to NIRSpec (which is the deepest) is as follows. Images through filters F090W+F444W are taken in parallel to exposures 1 and 2, F115W+F444W in parallel to exposures 3 and 4, F150W+F277W in parallel to exposure 5, and F200W+F356W in parallel to exposure 6.  For each filter, we use the DEEP8 readout pattern, with 7 groups and 1 integration per exposure, with 6 dithers in each exposure slot.  Thus each of the six exposure slots contain 8,245 seconds of imaging.  In this way, we will obtain $\sim 4.6$ hours each of F090W and F115W, $\sim 2.3$ hours in F150W, F200W, and F356W, and $\sim 9.2$ hours in F444W, reaching $5\sigma$ AB magnitudes limits for point-like sources in the range $\sim 29.2-29.7$ in each filter, according to the pre-launch ETC v.1.7 (Table~\ref{tab:parallels}).


The observational setup for the NIRCam imaging taken in parallel to NIRISS is as follows. For each direct image exposure (4 per grism angle per filter) in NIRISS we observe with one SW and one LW NIRCam filter in 6 groups of 311s each with the SHALLOW4 readout mode, and during each NIRISS grism exposure (4 exposures with small dithers) we observe with NIRCam for 6 groups totalling 4640s in the DEEP8 readout mode. In total, we observe with F444W for $\sim 6.5$ hours, F090W and F115W for a total of $\simeq 3.2$ hours each, F150W and F356W $\simeq 1.7$ hours each, and F200W and  F277W  for $\simeq 1.5$ hours each. Based on the pre-launch ETC v.1.7, with this strategy we expect to reach 5-$\sigma$ point source AB magnitude limits of $\sim29.0-29.5$ in each filter (Table~\ref{tab:parallels}).

The simulations presented below are meant to provide the reader with an intuition of the data quality to be expected.

\begin{figure*}
\includegraphics[width=0.5\linewidth]{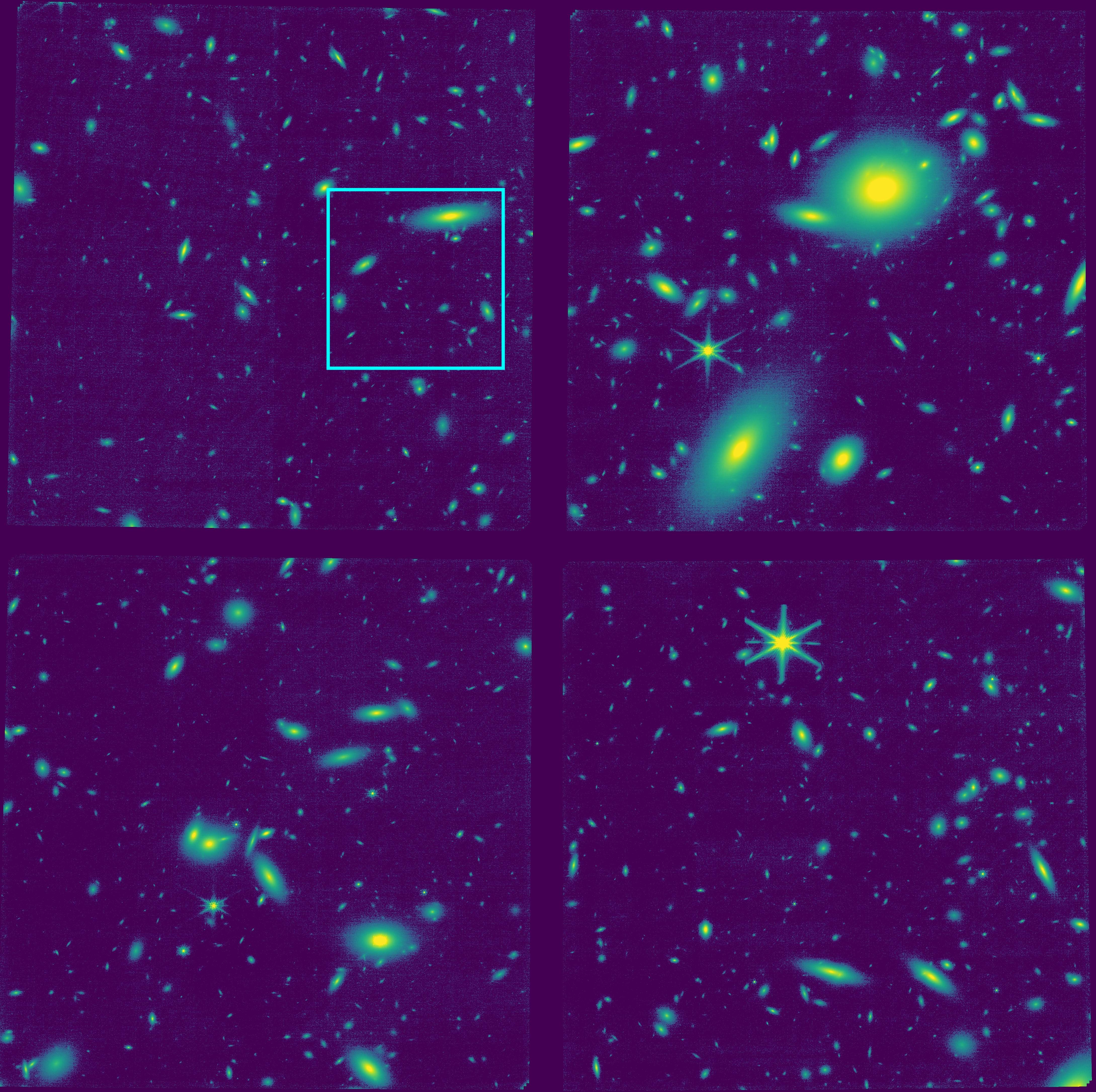}
\includegraphics[width=0.4975\linewidth]{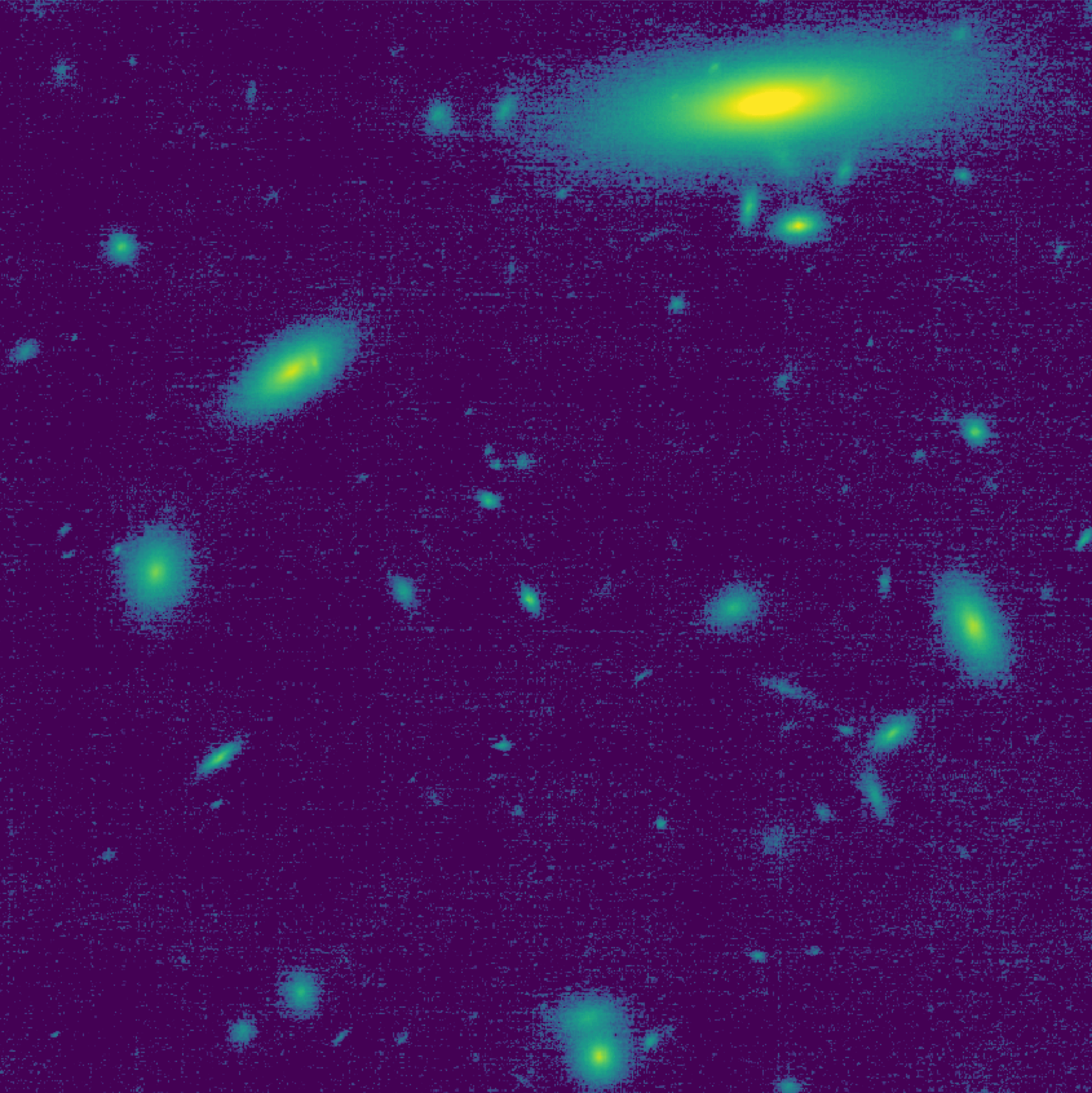}
\caption{Final simulated F150W mosaic generated by the NIRCam reduction pipeline. In the image on the left, a 4-quadrant mosaic (corresponding to a half FoV) is shown. On the right, a zoom-in of the region shown in the cyan box is shown. The faintest objects visible by eye on the images are fainter than $m_{F115W}\simeq 28.7$. }
\label{fig:nircam_sims}
\end{figure*}

\subsection{Simulated data sets}
\label{ssec:nircam_sims}


We performed extensive image reduction tests in preparation of the arrival of real data. We developed a simulation pipeline that, starting from the APT (Astronomer's Proposal Tool) file of the program, automatically creates the corresponding simulated NIRCam FoV. In practice, the simulation pipeline uses \textsc{EGG} \citep{schreiber17} to create a mock observed galaxy catalog centered on the celestial coordinates of the FoV. \textsc{EGG} exploits CANDELS data \citep{grogin11,koekemoer11} to empirically calibrate Monte-Carlo realizations of galaxy samples mimicking observed scaling relations and statistical properties; the code can be used to obtain simulated catalogues listing positions, magnitudes and morphological parameters (ellipticity, axis ratio, position angle) of galaxies observed with any set of filters in a given area of the sky and with any chosen depth. For a given APT configuration, we simulate a catalog that reaches about the $1\sigma$ depth of the final images, as estimated with the ETC, in order to include even object below the limiting depth. In the case of the F150W filter shown here  we extend the catalog to  limiting magnitude AB=31. We then add a stellar field, including the known bright stars in the FoV from the GAIA DR2 catalog, plus additional stars fainter than the GAIA limit using \textsc{TRILEGAL} \citep{girardi2005,girardi2012}, a populations synthesis code for simulating the stellar photometry of any desired Milky Way stellar field (the positions of the stars were then distributed randomly within the area, after removing those that are so bright that would be detected by GAIA). The colors of the GAIA stars are transformed into those expected in the given \jwst\ filter by assuming color transformations ($m_{JWST}=f(m_{BP}-m_{RP})+m_{RP}$) which has been derived using TRILEGAL. The galactic and stellar catalogs are then formatted to be used as input to \mirage\footnote{For the production of the NIRCam images presented here we have used the 2.2.1 version} (see the Appendix), to produce images with pixel scale $0\farcs031$. 

We then reduced the resulting images using the general STScI pipeline again as described in the Appendix. 
However, the current pipeline\footnote{For the reduction of the NIRCam images presented here we have used the 1.4.6 version} does not fully remove some horizontal and vertical striping patterns present in calibrated images. The striping is mainly due to the \textit{1/f} noise described by \citet{schlawin20}, and it is partially correlated at single amplifier scale. Four independent amplifier are, indeed, used to simultaneously read out four vertical stripes of 512 pixels in each detector. Therefore, starting from a solutions proposed by \citet{schlawin20}, we added a supplemental step to remove this noise from calibrated images. The step subtracts the median value from each row slide read by each amplifier, applying a sigma clipping to reduce the contribution of extended sources. This method is able to remove almost completely the striping. A downside of the algorithm is that it introduces artefacts in case of very extend sources (see Figure~\ref{fig:nircam_sims}). Alternative masking strategies are under study to avoid this unwanted side effect.


The \texttt{calwebb\_image3} pipeline is then run on the images to produce the final image and the relevant variance image. 
We have performed this exercise in all the seven filters, and we show in Figure~\ref{fig:nircam_sims} an example of the resulting scientific image F150W, computed in the case of the NIRSpec-parallel pointings. We note that, owing to the small dithering patterns adopted by the primary NIRSpec observations, the gaps betweeen detectors are very visible but cosmic ray and other defects are very effectively removed.

The simulations presented here are used to finalize the data reduction pipeline and prepare the analysis tools necessary to obtain the most accurate multi-wavelength galaxy catalogs and estimate their completeness and reliability. A quick analysis of the resulting completeness obtained by injecting in the image fake PSF-shaped point sources (with the expected FWHM of $0\farcs05$) of known magnitude, and using \textsc{SExtractor} \citep{bertin96} to detect them and measure their photometry shows that the completeness in the F150W at $m\sim 29$ is consistent with the ETC predictions, and recovered fluxes are similar to the input ones, although on average underestimated by approximately $0.1$ mags. 
We emphasize that these are preliminary results focused only at presenting to the general user the expected data quality and illustrate the tools necessary for the analysis of real data. Final recipes and more detailed tests on real data will be reported by the team after the data have acquired, calibrated, and reduced.

\subsection{Color selection and photometric redshifts}
\label{ssec:photoz}

We have adopted a 7-band filter strategy with the primary goal of allowing a robust selection of galaxies at $z>7$ via the Lyman-break technique and study their physical properties through a continuous sampling of the rest-frame optical emission. In particular, the F090W filter is essential to identify galaxies at $7<z<8.5$, while the F115W plays a similar role at $9<z<12$. However, the extended and continuous wavelength coverage and the high S/N of the resulting observations can deliver useful information for most of the observed galaxies, including those at lower $z$. 

To quantify the performance of the photometric redshift with the 7 \jwst\ bands of GLASS-JWST-ERS we have conducted the following simulations. We have taken a conservative approach to minimize the risks of obtaining an artificially high accuracy by using the same galaxy templates for simulating both the sample and recovering the redshift. We have built the input catalog using again the \textsc{EGG} simulator to predict a realistic distribution of colors and magnitudes for galaxies up to $z\sim 6$, augmented by a simulated catalog of Lyman-break galaxies computed using the BC03 \citep{BruzualCharlot2003} models, drawn from solar and sub-solar metallicity models with random age, exponentially increasing SFR, Calzetti attenuation curve \citep{calzetti2000} with $E(B-V)\leq 0.2$, Salpeter initial mass function (IMF) and IGM absorption from \citet{Inoue2014}.

\begin{figure*}
\includegraphics[width=0.99\linewidth, trim = 0cm 0cm 0cm 0cm]{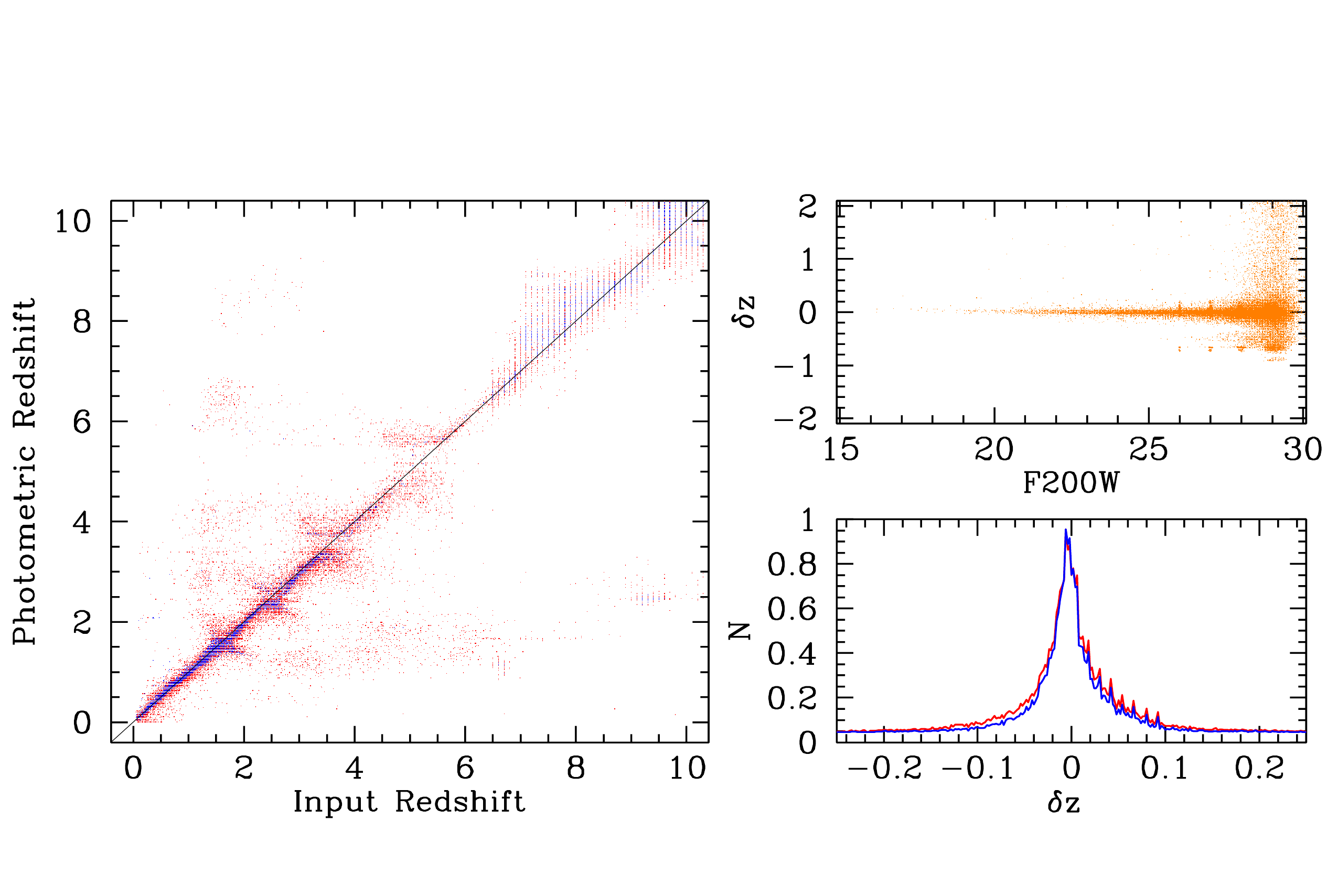}
\caption{\label{fig:parallels} Simulations predicting the accuracy of photometric redshifts with the GLASS-JWST-ERS NIRCam parallel filter set. {\it Left:} Comparison between the input redshift and the estimated photometric redshifts for objects at F200W$<28$ (blue points) and $28<$F200W$<29$ (red dots). {\it Right Upper:} Relative error $\delta z = (z_{phot}-z_{spec})/(1+z_{spec})$ as a function of the F200W observed magnitude.  {\it Right Lower}: Distribution of the relative error $\delta z$ for objects at F200W$<28$ (blue curve) and $28<$F200W$<29$ (red curve). See text for details of the simulations.}
\label{fig:photoz}
\end{figure*}

The resulting catalog of $\sim 30000$ sources has been perturbed with noise consistent with the  observations and then fitted with the {\it zphot} code \citep{Fontana2000, Merlin2021} adopting templates obtained from the Pegase 2 library \citep{Fioc2019}, with self-consistent treatment of the metallicity and dust evolution, a Rana-Basu IMF and the same IGM library. The resulting accuracy of photometric redshifts is shown in Figure~\ref{fig:photoz}. 

The \jwst-only photometric redshifts appear reliable and precise over a broad redshift range even though the filter set was optimized for $z>7$. This is not entirely surprising. At $z>1$, the \jwst\ filter set samples the entire nearUV-to-nearIR range, and is therefore able to track the position of the Balmer/$4000$\AA~ and other spectral breaks usually exploited by lower-z surveys. At lower $z$, it samples the position of the $1.6\mu$m peak of the stellar emission, which is also another strong redshift indicator \citep{Sawicki2002}. Overall, the resulting accuracy of $\delta z = \frac{z_{phot}-z_{spec}}{(1+z_{spec})}$ is $\sigma(\delta z)\sim 0.035$ at $m_{F200W}<29$. The fraction of outliers is negligible at $m_{F200W}<28$ and less than 5\% at $28<m_{F200W}<29$.  These numbers should be taken with a grain of salt, due to some simplifying assumptions such as for example not including emission lines), and the simulations should be repeated based on real data and a specific science goal.

However, in conclusion, our simulations show the GLASS-JWST-ERS NIRCam parallel observations will enable the identification of galaxies at virtually any redshifts in the range $z=0-10$ and provide unique information on their spectral energy distribution and rest frame optical size and morphology.

\section{Ancillary Magellan imaging}
\label{sec:Magellan}

In order to obtain imaging of the parallel fields at shorter wavelengths than NIRCam,
the field of A2744 was observed with MegaCam \citep{McLeod:2015} on the Magellan 2 Clay Telescope on September 7-8 2018. MegaCam field of view is large enough to ensure that the parallel fields will be covered irrespective of the final position angle of the primary observations.

Conditions were good with stable seeing $0\farcs5-0\farcs7$ throughout the run. Deep images were obtained through filters $g$ $r$ $i$, reaching $5$-$\sigma$ depths of 27.2, 26.3, 26.2\,AB, respectively, within $1''$ aperture.  The data were reduced using the MegaCam pipeline and custom scripts. An image is shown in the background of Figure~\ref{fig:FOV}. Reduced data and catalogs will be released together with the Stage 2 Science Enabling Products.

\section{Plans for release of high-level data product}
\label{sec:releases}

As for all the ERS programs, the data will be immediately public, with no proprietary period. However, the GLASS-JWST-ERS data set is multi-instrument and inherently complex. Therefore, we anticipate that much will be learned from this data set over an extended period of time.  To maximize the utility of our data set, we will proceed to release high level data products in two stages. Stage I will be a complete release of all data and science enabling products within six months of data acquisition.
Stage II will entail reprocessing all data based on lessons from Stage I and input from the community and our enlarged team. A second round of data/tools will be released as they are available. This will be completed within one year of data acquisition. 

Most data reduction will be done with the public \jwst\ pipeline. However, for NIRISS data we will compare the reduction using the \jwst\ pipeline with the reduction from publicly available code \texttt{grizli}\footnote{https://github.com/gbrammer/grizli}. Both reductions will be made available to the public. We will also provide a set of extracted NIRISS grism spectra from \texttt{grizli}. 

Availability and access to the data and high level data products will be announced through the GLASS-JWST-ERS website (\url{https://glass.astro.ucla.edu/ers/}). Users of the data are kindly requested to cite this paper as the source, in addition to the website.

\subsection{Stage I Science Enabling Products} 

Besides reduced data, we will release:
\begin{enumerate}
\item An object-based exploration tool. Interactive visualization of NIRISS grism data + imaging + NIRSpec spectroscopy will be critical to understanding these complex data sets. 
\item An intuitive RA/DEC-based NIRISS forced extraction tool. The default \grizli\ + NIRISS pipeline will extract spectra for sources detected in a pre-image. Yet, sometimes---e.g., in the case of faint, high-$z$ galaxies---forced-extraction of undetected sources at specific locations is needed. We will develop a tool that allows the user to easily perform this within the \grizli\ infrastructure, and visualize the results using the GUI.
\item Spectroscopic templates. Our program will be one of the first to obtain rest-frame optical spectra of $z>5$ sources. Such spectra have never been seen before. Instead, studies have had to rely on redshifting spectra of much lower-$z$ objects for forecasting/modeling. In order to enable the community to make more realistic forecasts and draw more accurate physical inferences, the core team will produce high $S/N$ template spectra by coadding NIRSpec data.
\item Catalogs of basic spectral quantities such as line fluxes will be made available for the NIRSpec targets. 
\item NIRCam-parallel catalogs. We will produce and release photometric catalogs
  of the parallel fields, focused on $z>7$ galaxies.
\end{enumerate}

\subsection{Stage II Science Enabling Products} 

The Stage~II release will include updates to the Stage~I products and a quantitative NIRISS/NIRSpec spectral comparison. As discussed above, these instruments provide each other with complementary spatial and spectral information. We will carry out a detailed comparison of the spectra obtained by two instruments for  our primary samples, with the goal of characterizing the effects of slit losses and limited spectral resolution in the inference of  emission line fluxes, star formation, dust extinction, and metallicity estimates.

The reduced MegaCam images of the field and associated catalogs will be publicly released together with the Stage II release.

\section{Summary}
\label{sec:summary}

In order to allow the community to make best use of the data, we have provided a comprehensive description of the GLASS-JWST-ERS program, which will obtain the deepest extragalactic data of the ERS campaign, by combining the power of \jwst\ with gravitational lensing magnification. 

The survey design was driven by the requirement to address two key scientific questions: i) What sources ionized the universe? ii) How do baryons cycle through galaxies? However, the instrumental setup and target selection are broad enough to enable a variety of investigations ranging from galaxy evolution in clusters, to the nature of dark, from high redshift passive galaxies and star clusters, to transient phenomena.

In the primary field, centered on cluster A2744, GLASS-JWST-ERS will obtain NIRISS slitless grism spectroscopy covering the wavelength range 1-2.2 $\mu$m. Simulations show that we will be able to achieve 50\% completeness for single emission lines down to fluxes of $\sim$10$^{-17}$ \ergscm, taking into account foreground contamination by cluster members and intracluster light. For magnified sources, fluxes as low as $\sim 10^{-18}$ \ergscm, will be detectable. By taking advantage of the orthogonally dispersed grisms we will be able to have at least one clean spectrum for $\sim$90\% of the sources. In terms of the key science drivers, the NIRISS spectra at $z>8$ will enable the measurement of the spatial distribution of \lya\  in comparison with the UV continuum, and the detection of UV lines such as \ion{C}{3} and \ion{C}{4} if present. At $z\sim2-3$ NIRISS will provide spatially resolved maps of star formation, dust extinction and gas metallicity, to study the cycle of baryons. NIRISS acquisition images will provide imaging of the primary field through filters F115W, F150W, F200W, supplementing the existing HFF images.

For $\sim$50 galaxies, we will obtain NIRspec multi object spectroscopy  covering the wavelength range 1-5$\mu$m, at resolution $R\sim2700$. As far as the key science drivers are concerned, at $z>8$ NIRspec will spectrally resolve \lya\ and provide its kinematics with respect to the systemic redshift traced by the rest frame optical lines, which will also measure the gas phase metallicity and dust extinction. At $z\sim2-5$ NIRspec will be able to measure the kinematics of the gas in starforming galaxies as well as a comprehensive set of UV and optical diagnostic lines for the characterization of star formation, gas enrichment, outflows and inflows.

For the sources in common between NIRISS and NIRspec,  the combination of spatially resolved low resolution grism spectroscopy from NIRISS and high spectral resolution slit spectroscopy from NIRSpec will enable  a quantitative comparison between the two spectrographs, as well as the combination of the benefits of spatial and spectral information.

In the two parallel fields, NIRCam will deliver 7-band imaging (spanning the wavelength range 0.8-5 $\mu$m) to a depth of 29-29.7 AB magnitudes over a combined area of 18 arcmin$^2$. The filter choice and depth is optimized for galaxies at $z>7$ and we expect to detect approximately 100-200 of them, providing the first constraints on their rest frame optical size and morphology, and new measurements of their abundance, stellar mass and star formation history.

All the raw data will be public immediately. High level data products will be delivered to the community in two stages. Stage I will take place within six months of data acquisition and will include in addition to the reduced data: an object based exploration tool; a forced extraction tool for NIRISS; spectroscopic templates for $z>5$ sources; catalogs of basic spectral quantities such as line fluxes; photometric catalogs from the NIRCam fields, optimized for $z>7$ galaxies. Stage II will take place within one year of data acquisition and will update all the stage one high level data products and add a quantitative comparison between NIRISS and NIRSpec for the targets in common.

\acknowledgments

Support for program JWST-ERS-1324 was provided by NASA through a grant from the Space Telescope   Science Institute, which is operated by the Association of Universities for Research in Astronomy, Inc., under NASA contract NAS 5-03127.

This paper includes data gathered with the 6.5 meter Magellan Telescopes located at Las Campanas Observatory, Chile

We are grateful to past members of the GLASS-JWST-ERS team for their contribution during the years (L.~Abramson, A.~Hoag, H.~Kuang, K.~Schmidt).

CAM acknowledges support by the VILLUM FONDEN under grant 37459 and the Danish National Research Foundation through grant DNRF140.

MB acknowledges support by the Slovenian national research agency ARRS through grant N1-0238.

We acknowledge financial support through grant PRIN-MIUR 2017WSCC32 ``Zooming into dark matter and proto-galaxies with massive lensing clusters''.

GBC acknowledges the Max Planck Society for financial support through the Max Planck Research Group for S. H. Suyu and the academic support from the German Centre for Cosmological Lensing. 

MN acknowledges support from grant INAF-1.05.01.86.20

This work utilizes gravitational lensing models produced by PIs Bradac, Natarajan \& Kneib (CATS), Merten \& Zitrin, Sharon, Williams, Keeton, Bernstein and Diego, and the GLAFIC group. This lens modeling was partially funded by the \hst\ Frontier Fields program conducted by STScI. STScI is operated by the Association of Universities for Research in Astronomy, Inc. under NASA contract NAS 5-26555. The lens models were obtained from the Mikulski Archive for Space Telescopes (MAST).

This work was partially supported by the Australian Research Council Centre of Excellence for All-Sky Astrophysics in 3 Dimensions (ASTRO-3D) through award CE170100013.  BMP and BV acknowledge the funding received from the European Research Council (ERC) under the European Union's Horizon 2020 research and innovation programme (grant agreement No. 833824).

KG, TN and CJ acknowledge support from Australian Research Council Laureate Fellowship FL180100060.

LY acknowledges support by JSPS KAKENHI Grant Number JP 21F21325.

\appendix
\section{Simulation Tools \& Data Reduction Pipeline}
\label{sec:tools}

Throughout various sections in the paper we detail efforts to construct realistic simulations of NIRISS and NIRCam data sets (imaging and spectroscopic). Given the frameworks for the construction of these simulated data sets make use of the same software tools, we provide here a brief description of the main tools and procedures used. In short, both NIRISS and NIRCam images are constructed via the Multi-Instrument RAmp GEnerator (\mirage\footnote{\url{https://mirage-data-simulator.readthedocs.io/en/latest/}}) code and reduced with the official \jwst\ data reduction pipeline.

\mirage\ is a Python package developed by the STScI NIRCam and NIRISS instrument teams to simulate realistic data for \jwst's NIRCam, NIRISS, and FGS instruments. It has been designed to create a single simulated exposure from a single input file, given in \texttt{yaml} format, which contains all the configuration to simulate the data. Making use of the APT, the XML and the pointing files needed by \mirage\ as input - together with constructed source catalogs (of point sources, galaxies and stars) - are obtained, and all of the \texttt{yaml} files associated with the ERS observing program are generated; one can then generate with \mirage\ each of the NIRISS or NIRCam ramp images associated with the ERS program, with an output format that is standardized to the real \jwst\ data format, allowing to process the simulated images with the official \jwst\ imaging data reduction pipeline.

The official \jwst\ pipeline\footnote{\url{https://jwst-pipeline.readthedocs.io/en/latest/jwst/introduction.html}} is also provided as Python library, where the application of three subsequent pipelines is required to calibrate NIRISS and NIRCam images. The first of these, \texttt{calwebb\_detector1}, is applied to ramp images from both instruments and to all kinds (direct or spectroscopic), correcting for detector-level artefacts and converting to countrate images. Subsequent steps then differentiate between imaging (\texttt{calwebb\_image2}) and spectroscopic output (\texttt{calwebb\_spec2}), however largely achieve the same goal: applying astrometric and photometric calibrations to obtain fully reduced  and calibrated images (in MJy/sr). A third pipeline, \texttt{calwebb\_image3}, takes care of (for direct imaging only) the relative alignment among different frames, the creation of a stacked image, and the extraction of a source catalogue. Unless otherwise stated, both the NIRISS and NIRCam simulations described in this paper adopt the above \mirage\ and pipeline procedures.

\end{document}